\documentclass[]{aastex631}

\usepackage{color}

\usepackage[T1]{fontenc}
\usepackage{graphicx}	
\usepackage{amsmath}	
\usepackage{amssymb}	
\usepackage{threeparttable}
\usepackage{booktabs}
\usepackage{makecell}
\usepackage{enumerate}
\usepackage[shortlabels]{enumitem}
\usepackage{color}
\usepackage{multirow}  
\DeclareSymbolFont{CMletters}{OT1}{cmtt}{m}{ui}
\DeclareMathSymbol{g}{\mathord}{CMletters}{`g}
\usepackage{newtxmath}

\makeatother

\newcommand{\nmol}{33 }
\newcommand{\niso}{58 }
\newcommand{\nexomols}{$129\,237\,384$ }
\newcommand{\nexomolt}{$727\,841\,563\,143$ }
\newcommand{\nhrs}{$238\,263$ }
\newcommand{\nhrl}{$24\,307\,135$ }
\newcommand{\MARVEL}{{\textsc{MARVEL} }}
\newcommand{\ExoMol}{{\textsc{ExoMol} }}
\newcommand{\ExoMolHR}{{\textsc{ExoMolHR} }}
\newcommand{\PyExoCross}{{\textsc{PyExoCross} }}
\usepackage{amssymb}

\graphicspath{{./}}

\begin{document}

\title{\textsc{ExoMolHR}: A Relational Database of Empirical High-Resolution Molecular Spectra}

\correspondingauthor{Jingxin Zhang}
\email{jingxin.zhang.19@ucl.ac.uk}

\author[0000-0002-6374-6251]{Jingxin Zhang}
\affiliation{Department of Physics and Astronomy, University College London \\
Gower Street, WC1E 6BT London, UK}

\author[0000-0001-6604-0126]{Christian Hill}
\affiliation{International Atomic Energy Agency, Vienna A-1400, Austria}

\author[0000-0002-4994-5238]{Jonathan Tennyson}
\affiliation{Department of Physics and Astronomy, University College London \\
Gower Street, WC1E 6BT London, UK}

\author[0000-0001-9286-9501]{Sergei N. Yurchenko}
\affiliation{Department of Physics and Astronomy, University College London \\
Gower Street, WC1E 6BT London, UK}



\begin{abstract}

\textsc{ExoMolHR} is an empirical, high-resolution molecular spectrum calculator for the high-temperature molecular line lists available from the \textsc{ExoMol} molecular database. 
Uncertainties, where available, in recommended \textsc{ExoMol} datasets are used to select highly accurate spectral lines. These lines largely rely on empirical energy levels generated through the MARVEL (measured active rotation vibration energy levels) procedure, which is being systematically used to improve the energy and transition data provided by the \textsc{ExoMol} database. The freely accessible \textsc{ExoMolHR} database provides line positions with  calculated intensities for a user-specified  wavenumber/wavelength range and temperature. Spectra can be plotted on the \textsc{ExoMolHR} website \url{https://www.exomol.com/exomolhr/} or downloaded as a csv file. Cross sections can be calculated using the Python program \textsc{PyExoCross}. 
The \textsc{ExoMolHR} database currently provides \nhrl spectral lines for \nmol molecules and \niso isotopologues; these numbers will increase as the \textsc{ExoMol} database is updated.

\end{abstract}



\section{Introduction} \label{sec:intro}

The assignment and analysis of astronomical spectra is generally performed using laboratory
spectra. A number of spectroscopic databases are available to aid astronomers in this activity.
Notable for long wavelength studies are the Cologne Database for Molecular Spectroscopy
(CDMS) \citep{CDMS} and the Jet Propulsion Laboratory (JPL) submillimeter, millimeter, and microwave spectral line catalog \citep{JPL} which have proved to be the mainstay of observational radio astronomy. Similarly, there are careful compilations of atomic spectra provided by
sources such as the National Institute of Standards and Technology (NIST) Atomic Databases \citep{NIST,NIST_ASD}, \citet{11Kurucz.db} and the  Vienna Atomic Line Database (VALD) \citep{VALD3}. For molecular spectra at shorter wavelengths the situation is less
straightforward. The HITRAN \citep{jt857} database contains line data at largely infrared wavelengths for 55 molecules selected primarily for studies of the Earth's atmosphere at
temperatures close to 296 K.
Similarly GEISA (Gestion et Etude des Informations Spectroscopiques Atmosph\'eriques) database
contains line data for 57 molecules which are substantially the same as those given by HITRAN.
We note that the HITEMP project, while extending the temperature range of HITRAN data,
only currently contains line lists for 8 molecules \citep{jt480,jt763}, all of which
feature in both HITRAN and GEISA.

\citet{jt528} started the \textsc{ExoMol} database to provide molecular line lists for exoplanet and other (hot) atmospheres. Originally the emphasis of this project has been on completeness of these
line lists \citep{jt572} rather than accuracy. However, over the last few years we have
increasingly used experimental data to improve the accuracy of line positions provided
by the database, see \citet{jt948} for example. As a result the latest \textsc{ExoMol} data release
\citep{jt939} claims 41 molecules for which high accuracy transition wavenumbers are
available for at least one isotopologue. The molecules considered in the \textsc{ExoMol} database are selected for their astronomical
importance and data are provided for studies covering extended temperature ranges. The provision of
high accuracy data in \textsc{ExoMol} is an important step forward; however, the \textsc{ExoMol} database is
huge, much larger than any of those quoted above, as it contains in the region of 10$^{13}$
transitions. As described below, the high accuracy lines given by \textsc{ExoMol} form only a small a proportion of the total, currently \nhrl transitions. 
The sheer volume of data in the \textsc{ExoMol} database makes it unsuitable for the sort of detailed spectroscopic
analysis involved in, for example, initial line assignments. The aim of this work is to
provide easy access to the high resolution molecular data contained in the \textsc{ExoMol} data
base which can be used for assignment and analysis of strong lines.

\textsc{ExoMol} has been undertaking a systematic activity improving the accuracy of the energy levels, and hence 
transition frequencies, provided using the MARVEL (measured active rotation vibration energy levels) procedure
\citep{jt412,jt908}. A number of dedicated MARVEL studies have been performed recently on species such as C$_2$ \citep{jt809}, AlO \citep{jt835}, H$_3$$^+$ and its
deuterated isototopologues \citep{jt890}, formaldehyde \citep{jt828,jt906}, methane \citep{jt926} and VO \citep{jt869}; recent
work by \citet{jt948} has shown how MARVEL studies can be leveraged to extend the number of accurately determined
energy levels and transitions. Accurate energy levels generated by MARVEL and related techniques are 
used by \textsc{ExoMol} to improve the accuracy of their data by replacing calculated energy levels with empirically determined ones.
It is important to note that this technique usually leads that many more  high accuracy transitions being provided by \textsc{ExoMolHR} than have actually  been observed in the laboratory, see \citet{jt828} or  \citet{jt869} for example.

\section{The \textsc{ExoMol} Database} \label{sec:background}

The \textsc{ExoMolHR} database stores largely empirical high-resolution spectral lines extracted from the \textsc{ExoMol} database. While line strength is not used in the extraction procedure,  these lines are largely the  stronger ones.  Before introducing \textsc{ExoMolHR}
we summarize the contents of the \textsc{ExoMol} database.

The \textsc{ExoMol} project was designed for interpreting spectra and modelling atmospheres of (hot) exoplanets, cool stars, brown dwarfs, and other hot astronomical objects by providing comprehensive lists of high-temperature molecular spectroscopic transitions \citep{jt631,jt810}. In practice,  \textsc{ExoMol} data is also appropriate for non-astronomical applications such as   combustion studies.
The \textsc{ExoMol} database is accessible from the \textsc{ExoMol} website \url{https://www.exomol.com/}. The current \textsc{ExoMol} database 2024 release provided spectroscopic data for 91 molecules and 224 isotopologues \citep{jt939}. While the majority of these datasets (55 molecules) were generated in-house by members of the
\textsc{ExoMol} project, the \textsc{ExoMol} database also hosts datasets for 36 molecules generated either
in house at University College London (UCL) for other reasons or externally. Many of these hosted datasets contain high accuracy transition data for the systems concerned, for example the datasets from Bernath's MolLIST project \citep{MOLLIST} are based on laboratory high resolution measurements. However, \textsc{ExoMolHR} requires explicit uncertainties to be present in the
data to enable their inclusion. At present explicit uncertainties are only available for three external datasets generated by McKemmish's group from UNSW (University of New South Wales) namely
CN    \citep{21SyMcXx.CN}, ZrO \cite{23PeTaMc.ZrO} and
NH \cite{24PeMcxx.NH}.

The \ExoMol database has a well-defined data structure \citep{jt548,jt939}.
The  \ExoMol master file (\texttt{exomol.json})  points to a definition file for each line list with the version number given
by the latest update date in the format \texttt{YYYYMMDD}. The
\ExoMol database provides both native format \texttt{.def} and  JSON (JavaScript Object Notation, \citet{JSON}) format \texttt{.def.json} (\texttt{<ISOTOPOLOGUE>\_\_<DATASET>.def.json}) files which provide the information on molecular characteristics, uncertainty availability, quantum number labels and formats and much else to allow for intepretation and automated processing the states and transitions. We used the information provided by these .def files to generate the \ExoMolHR database.

The \textsc{ExoMol} database is structured around two core file types: states (\texttt{.states}) and transitions (\texttt{.trans}) file \citep{jt548}. The \texttt{.states} files document the energy levels and their associated quantum numbers, while the \texttt{.trans} files store the Einstein $A$ coefficients identified by upper and lower state indices that reference the corresponding entries in the \texttt{.states} files. Due to their potentially large size, especially the \texttt{.trans} files, both types of files are compressed using the \texttt{.bz2} format. For some large line lists, the transitions are divided into multiple files in a series, each covering a specific wavenumber region as indicated in the file name. Partition function (\texttt{.pf}) files provided by the \textsc{ExoMol} database are also used  by the \textsc{ExoMolHR} database to calculate transition intensities. 
Tables~\ref{tab:states}, \ref{tab:trans}, and \ref{tab:pf} illustrate the structure of the \texttt{.states}, \texttt{.trans}, and \texttt{.pf} files, respectively. Note that the uncertainty column, labeled $\Delta E$ in Table~\ref{tab:states}, is a relatively new feature in the  \textsc{ExoMol} database and not all datasets contain uncertainties.  \textsc{ExoMolHR} requires the uncertainties to be present so that datasets without an $\Delta E$ column (as determined from the  \textsc{ExoMol} \texttt{.def.json} file) are simply ignored. However, these datasets are substantially ones that have not been updated using \MARVEL energy levels which therefore do not contain any high accuracy data of interest for  \textsc{ExoMolHR}.
 
\begin{deluxetable}{llll}
\tabletypesize{\small}
\tablecolumns{4} 
\setlength{\tabcolsep}{2.9mm}{
\tablecaption{Specification of the \texttt{.states} file including extra data options; the formats at the end of the table are for the compulsory section only \citep{jt939}. \label{tab:states}}
\tablehead{
\colhead{Field\ \ \ \ } & \colhead{Fortran Format\ } & \colhead{C Format\ \ \ \ \ } & \colhead{Description\ \ \ \ \ \ \ \ \ \ \ \ \ \ \ \ \ \ \ \ \ \ \ \ \ \ \ \ \ \ \ \ \ \ \ \ \ \ \ \ \ \ \ \ \ \ \ \ \ \ \ \ \ \ \ \ \ \ \ \ \ \ \ \ \ \ \ \ \ \ \ \ \ \ \ \ \ \ \ \ \ \ \ }
}
\startdata
$i$              & \texttt{I12}           & \texttt{\%12d}         & State ID \\
$\tilde{E}$      & \texttt{F12.6}         & \texttt{\%12.6f}       & Recommended state energy in $\mathrm{cm^{-1}}$ \\
$g_\mathrm{tot}$ & \texttt{I6}            & \texttt{\%6d}          & Total state degeneracy \\
$J$              & \texttt{I7/F7.1}       & \texttt{\%7d/\%7.1f}   & Total angular momentum quantum number, $J$ or $F$ (integer/half-integer ) \\
$\Delta E$       & \texttt{F12.6}         & \texttt{\%12.6f}       & Uncertainty in the state energy in $\mathrm{cm^{-1}}$ \\
$\tau$           & \texttt{ES12.4}        & \texttt{\%12.4E}       & State lifetime (aggregated radiative and predissociative lifetimes) in s \\
($g$)            & \texttt{F10.6}         & \texttt{\%10.6f}       & Land\'e $g$-factor (optional) \\
(QN)             & See \texttt{.def} file & See \texttt{.def} file & State quantum numbers, may be several columns (optional) \\
(Abbr)           & \texttt{A2}            & \texttt{\%2s}          & Abbreviation giving source of state energy (optional) \\
($\tilde{E_0}$)  & \texttt{F12.6}         & \texttt{\%12.6f}       & Calculated state energy in $\mathrm{cm^{-1}}$ (optional)
\enddata
}
\end{deluxetable}

\begin{deluxetable}{llll}
\tabletypesize{\small}
\tablecolumns{4} 
\setlength{\tabcolsep}{1.8mm}{
\tablecaption{Specification of the \texttt{.trans} file including extra data options. \label{tab:trans}}
\tablehead{
\colhead{Field\ \ \ \ } & \colhead{Fortran Format\ } & \colhead{C Format\ \ \ \ \ } & \colhead{Description\ \ \ \ \ \ \ \ \ \ \ \ \ \ \ \ \ \ \ \ \ \ \ \ \ \ \ \ \ \ \ \ \ \ \ \ \ \ \ \ \ \ }
}
\startdata
$f$ & \texttt{I12} & \texttt{\%12d} & Upper state ID \\
$i$ & \texttt{I12} & \texttt{\%12d} & Lower state ID \\
$A$ & \texttt{ES10.4} & \texttt{\%10.4E} & Einstein $A$ coefficient in $\mathrm{s^{-1}}$ \\
$\tilde{\nu}_{fi}$ & \texttt{E15.6} & \texttt{\%15.6E} & Transition wavenumber in cm$^{-1}$ (optional)
\enddata
}
\end{deluxetable}

\begin{deluxetable}{llll}
\tabletypesize{\small}
\tablecolumns{4} 
\setlength{\tabcolsep}{1.8mm}{
\tablecaption{Specification of the \texttt{.pf} partition function file format. \label{tab:pf}}
\tablehead{
\colhead{Field\ \ \ \ } & \colhead{Fortran Format\ } & \colhead{C Format\ \ \ \ \ } & \colhead{Description\ \ \ \ \ \ \ \ \ \ \ \ \ \ \ \ \ \ \ \ \ \ \ \ \ \ \ \ \ \ \ \ \ \ \ \ \ \ \ \ \ \ }
}
\startdata
$\textsl{T}$ & \texttt{F8.1} & \texttt{\%8.1f} & Temperature in K \\
$\textsl{Q}(T)$ & \texttt{F15.4} & \texttt{\%15.4f} & Partition function (dimensionless)
\enddata
}
\end{deluxetable}

\section{Methodology and algorithm} \label{sec:methodology}

\begin{figure}[ht!]
\epsscale{1.18}
\plotone{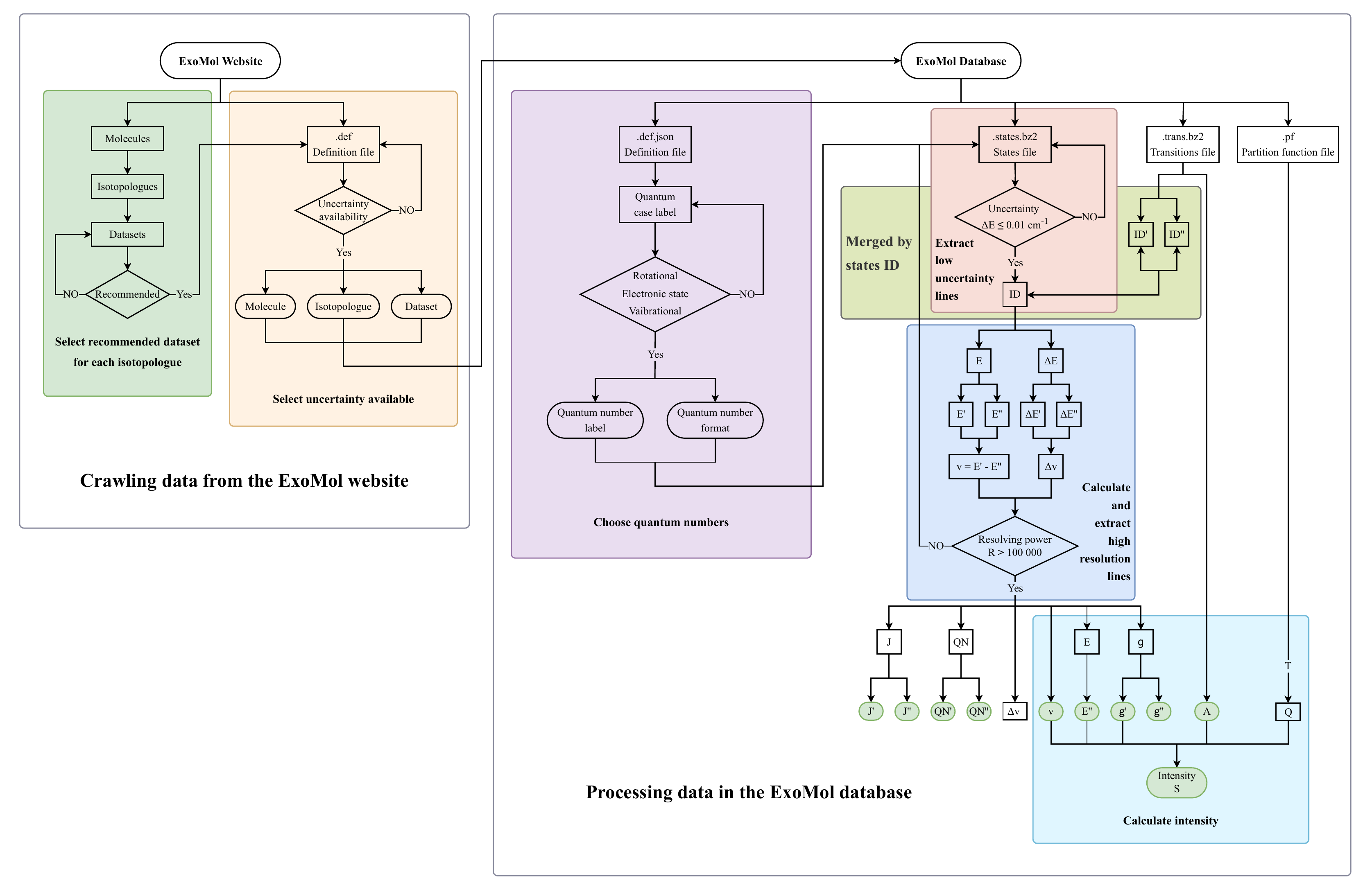}
\caption{Process flowchart for \ExoMolHR database generation. 
\label{fig:hr process}}
\end{figure}

The \ExoMolHR database is designed to provide highly accurate line positions for high-resolution studies, such as identifying spectral lines or simulating high-resolution spectra. 
Figure~\ref{fig:hr process} illustrates how the \ExoMolHR database is generated. This process requires data to be obtained separately from the \textsc{ExoMol} websites and its database by following steps.
Initially, we scraped data from the \textsc{ExoMol} website for all molecules and  isotopologues. We extract the recommended dataset for each isotopologue and then iterate through their definition file (\texttt{.def.json}), processing only those datasets for which explicit uncertainties are available.
Next, we process the line list files to obtain the labels and formats of the quantum numbers from the \textsc{ExoMol} JSON format definition files (\texttt{.def.json}) and keep only the vibrational, rotational, and electronic state quantum numbers in the \texttt{.states} files. 
Subsequently, we extract those lines from the \texttt{.states} and \texttt{.trans} files of the \textsc{ExoMol} database which are determined to high accuracy (low uncertainty); in practice energy levels with an uncertainty $\Delta E \le$ 0.01 cm$^{-1}$ in the \texttt{.states} file are retained for processing into transition wavenumbers. 

Resolving power, $R$,  is defined by the wavelength $\lambda$ and its uncertainty $\Delta \lambda$. Here we use the transition wavenumber $\tilde{\nu}$ and its uncertainty $\Delta \tilde{\nu}$ to calculate the resolving power $R$ as 
\begin{equation}
\label{eq:resolution}
    R = \frac{\lambda}{\Delta \lambda} \approx \frac{\tilde{\nu}}{\Delta \tilde{\nu}},
\end{equation}
where the transition wavenumber (frequency) $\tilde v$ is equal to the upper state energy $\tilde E'$ minus the lower state energy $\tilde E''$:
\begin{equation}
\label{eq:wavenumber}
    \tilde{\nu} = \tilde E' - \tilde E'',
\end{equation}
and 
\begin{equation}
    \Delta \tilde{\nu} = \sqrt{\left (\Delta E^{'}\right )^2 + \left (\Delta E^{''}\right )^2}.
\end{equation}
$\Delta E^{'}$ and $\Delta E^{''}$ are the uncertainties of the upper and lower state energy levels, respectively.
Finally, we scrape high resolution transitions which are determined to correspond to $R>100\,000$.
The \ExoMolHR database supports calculating absorption line intensities $I_{fi}$ (cm molecule$^{-1}$) at different temperatures $T$ in K provided by users with the equation (\ref{eq:absintensity}) \citep{96JoLaIw.CH,jt708}:
\begin{equation} 
\label{eq:absintensity}
    I_{f \gets i} = \frac{g'_f {A}_{fi}}{8 \pi c \tilde{\nu}^2_{fi}} \frac{e^{-c_2 \tilde{E}''_i / T} (1 - e^{-c_2 \tilde{\nu}_{fi} / T })}{\textsl{Q}(T)}.
\end{equation}
$g'_f$ is the upper state degeneracy and $\tilde{E}''_i$ is the lower state energy in \texttt{.states} file. $A_{fi}$ is the Einstein $A$-coefficient (s$^{-1}$) in \texttt{.trans} file. $\tilde{\nu}_{fi}$ is the transition wavenumber (cm$^{-1}$) in \texttt{.trans} file for small datasets or calculated by equation (\ref{eq:wavenumber}). $c_2= hc/k_B$ is the second radiation constant (cm K), where $h$ is the Planck constant (erg s), $c$ is the speed of light (cm s$^{-1}$), and $k_B$ is the Boltzmann constant (erg K$^{-1}$). $\textsl{Q}(T)$ is the temperature-dependent partition function given by the \textsc{ExoMol} partition function file (\texttt{.pf}).

\section{\textsc{ExoMolHR} Database Overview} \label{sec:database}

\ExoMolHR is an extensive high-resolution molecular spectroscopic database based on the \textsc{ExoMol} database which contains  molecular line lists  designed for studies of  hot environments. 
\ExoMolHR database is hosted at \url{https://www.exomol.com/exomolhr/}. This is initial offering contains lines for \niso\ isotopologues associated with \nmol\  molecules which are supported on the \ExoMolHR website, see Table~\ref{tab:HRlist} below. 

This section will introduce the file structure and features (Section~\ref{sec:filestructure}), use cases (Section~\ref{sec:usecases}), the application programming interface (API) (Section~\ref{sec:API}), and a summary of line list in the \ExoMolHR database.

\subsection{File structure and features} \label{sec:filestructure}

Table~\ref{tab:csvfmt} specifies the \ExoMolHR database file structure and Table~\ref{tab:H2O} is a sample of the \ExoMolHR database file (\texttt{.csv}) for $^1$H$_2$$^{16}$O at temperature $T=296$ K. Transitions form the core of the \ExoMolHR database including the frequency (which can be extracted for a  user-specified wavenumber range), transition uncertainty, Einstein $A$-coefficient, intensity (at a user-specified temperature), lower state energy level, total degeneracy, angular momentum, and quantum numbers for upper and lower states. The output \texttt{.csv} are not provided on the website but their contents (with an extra intensity column for the user specified temperature) can be extracted simply by requesting data for the given wavenumber or wavelength range for the species of interest. 

\texttt{ExoMol} state files, particularly  those for polyatomic molecules, often have multiple sets of quantum numbers.
However, for simplicity, the \ExoMolHR database only provides a single set of quantum numbers for each state which are generally based on a standard, normal mode description of the vibrational motion. If needed, the alternative
quantum number descriptions can be obtained from the appropriate \texttt{.states} file.

\begin{deluxetable}{llll}
\tabletypesize{\small}
\tablecolumns{4} 
\setlength{\tabcolsep}{1.7mm}{
\tablecaption{Specification of the output \texttt{.csv} \textsc{ExoMolHR} file format. \label{tab:csvfmt}}
\tablehead{
\colhead{Field\ \ \ } & \colhead{Fortran Format\ } & \colhead{C Format\ \ \ \ \ } & \colhead{Description\ \ \ \ \ \ \ \ \ \ \ \ \ \ \ \ \ \ \ \ \ \ \ \ \ \ \ \ \ \ \ \ \ \ \ \ \ \ \ \ \ \ \ \ \ \ \ \ \ \ \ \ \ \ \ \ \ \ \ \ \ \ \ \ \ \ \ \ \ \ \ \ \ \ \ \ \ \ \ \ \ \ \ \ \ \ \ \ \ \ \ \ \ }
}
\startdata
$\tilde{\nu}$ or $\tilde{\lambda}$ & \texttt{F12.6} & \texttt{\%12.6f} & Transition wavenumber/frequency in cm$^{-1}$ or wavelength in $\mu$m \\
$\Delta \nu$    & \texttt{F12.6}         & \texttt{\%12.6f}       & Unceratinty in the transition wavenumber in cm$^{-1}$ \\
$A$           & \texttt{ES10.4}        & \texttt{\%10.4E}       &  Einstein $A$ coefficient in s$^{-1}$ \\
$S$           & \texttt{ES10.4}        & \texttt{\%10.4E}       &  Absorption intensity at the user specified temperature in cm$/$molecule \\
$E''$         & \texttt{F12.6}         & \texttt{\%12.6f}       & Lower state energy in cm$^{-1}$ \\
$g'$          & \texttt{I6}            & \texttt{\%6d}          & Total state degeneracy for upper state \\
$g''$         & \texttt{I6}            & \texttt{\%6d}          & Total state degeneracy for lower state \\
$J'$          & \texttt{I7/F7.1}       & \texttt{\%7d/\%7.1f}   & Total angular momentum quantum number (integer/half-integer) for upper state \\
$J''$         & \texttt{I7/F7.1}       & \texttt{\%7d/\%7.1f}   & Total angular momentum quantum number (integer/half-integer) for lower state \\
(QN$'$)       & See \texttt{.def} file & See \texttt{.def} file & Quantum number for upper state (optional) \\
(QN$''$)      & See \texttt{.def} file & See \texttt{.def} file & Quantum number for lower state (optional)\\
\enddata
}
\tablecomments{
Quantum number labels and formats are defined in the \textsc{ExoMol} definition file, \texttt{.def.json}.
}
\end{deluxetable}
\begin{deluxetable}{ccccccccccccccccccccc}
\tabletypesize{\footnotesize}
\tablecolumns{21} 
\setlength{\tabcolsep}{1.2mm}{
\tablecaption{Extract from the \textsc{ExoMolHR} file of $^1$H$_2$$^{16}$O at temperature $T=296$ K (\texttt{yyyymmddhhmmss\_\_1H2-16O\_\_296K.csv}). \label{tab:H2O}}
\tablehead{
\colhead{$\tilde{\nu}$} & \colhead{$\Delta \nu$} & \colhead{$A$} & \colhead{$S$} & \colhead{$\tilde{E}''$} & \colhead{$g'$} & \colhead{$g''$} & \colhead{$J'$} & \colhead{$J''$} & \colhead{$\Gamma_{\textrm{rve}}'$} & \colhead{$v'_1$} & \colhead{$v'_2$} & \colhead{$v'_3$} & \colhead{$K'_a$} & \colhead{$K'_c$} & \colhead{$\Gamma''_{\textrm{rve}}$} & \colhead{$v''_1$} & \colhead{$v''_2$} & \colhead{$v''_3$} & \colhead{$K''_a$} & \colhead{$K''_c$} 
}
\startdata
500.018142 & 0.000029 & 2.3792E$-$03 & 1.3746E$-$30 & 4195.818039 & 15 & 13 & \ 7 & \ 6 & A2 & 1 & 0 & 0 & \ 5 & \ 3 & A1 & 0 & 0 & 1 & 1 & \ 6 \\
500.035515 & 0.000009 & 1.7621E$-$01 & 2.3697E$-$24 & 2248.063135 & 27 & 27 & 13  & 13  & A1 & 0 & 0 & 0 & \ 6 & \ 8 & A2 & 0 & 0 & 0 & 3 & 11  \\
500.436638 & 0.000023 & 3.8196E$-$04 & 1.3245E$-$30 & 4052.836634 & 45 & 45 & \ 7 & \ 7 & B1 & 0 & 0 & 1 & \ 3 & \ 5 & B2 & 0 & 2 & 0 & 3 & \ 4 \\
500.621556 & 0.000028 & 9.2079E$-$02 & 3.4260E$-$23 & 1774.750359 & 75 & 75 & 12  & 12  & B2 & 0 & 0 & 0 & \ 5 & \ 8 & B1 & 0 & 0 & 0 & 2 & 11  \\
500.661272 & 0.000012 & 2.6278E$-$04 & 2.3620E$-$30 & 4006.071011 & 93 & 99 & 15  & 16  & B1 & 0 & 1 & 0 & \ 2 & 13  & B2 & 0 & 0 & 0 & 7 & 10  \\
500.878899 & 0.000050 & 1.9196E$-$01 & 1.0828E$-$24 & 2426.195175 & 27 & 27 & 13  & 13  & A2 & 0 & 0 & 0 & \ 7 & \ 7 & A1 & 0 & 0 & 0 & 4 & 10  \\
501.573101 & 0.000071 & 5.8196E$+$01 & 2.2409E$-$22 & 2471.254912 & 23 & 21 & 11  & 10  & A1 & 0 & 0 & 0 & 10  & \ 2 & A2 & 0 & 0 & 0 & 9 & \ 1 \\
501.573504 & 0.000071 & 5.8196E$+$01 & 6.7226E$-$22 & 2471.254532 & 69 & 63 & 11  & 10  & B1 & 0 & 0 & 0 & 10  & \ 1 & B2 & 0 & 0 & 0 & 9 & \ 2 \\
502.064969 & 0.000043 & 6.5476E$-$03 & 2.4083E$-$30 & 4257.786653 & 13 & 11 & \ 6 & \ 5 & A2 & 0 & 0 & 1 & \ 6 & \ 1 & A1 & 1 & 0 & 0 & 4 & \ 2 
\enddata
}
\tablecomments{
yyyymmddhhmmss: Filename is structured with 14 digits of created date and time, iso-slug, and temperature. yyyymmdd means date with year, month, and day. hhmmss means time with hour, minute, and second; \\
$\tilde{\nu}$: Transition wavenumber/frequency in cm$^{-1}$; \\
$\Delta \nu$: Unceratinty in the transition wavenumber in cm$^{-1}$; \\
$A$: Einstein $A$ coefficient in s$^{-1}$; \\
$S$: Absorption intensity in cm$/$molecule; \\
$\tilde{E}''$: Lower state energy in cm$^{-1}$; \\
$g'$ and $g''$: Total state degeneracy for upper and lower state; \\
$J'$ and $J''$: Total angular momentum rotational quantum number for upper and lower state, excluding nuclear spin; \\
$\Gamma_{\textrm{rve}}'$ and $\Gamma_{\textrm{rve}}''$: Rovibrational symmetry label for upper and lower state; \\
$v'_1$ and $v''_1$: $v_1$ symmetric stretch quantum number for upper and lower state; \\
$v'_2$ and $v''_2$: $v_2$ bend quantum number for upper and lower state; \\
$v'_3$ and $v''_3$: $v_3$ asymmetric stretch quantum number for upper and lower state; \\
$K'_a$ and $K''_a$: $K_a$ rotational quantum number for upper and lower state; \\
$K'_c$ and $K''_c$: $K_c$ rotational quantum number for upper and lower state. 
}
\end{deluxetable}

The \ExoMolHR files are named in format \texttt{<yyyymmddhhmmss>\_\_<ISO-SLUG>\_\_<T>K.csv} where \texttt{<yyyymmddhhmmss>} is a 14-digit character string corresponding to a timestamp in the format of year (\texttt{yyyy}), month (\texttt{mm}), day (\texttt{dd}), hour (\texttt{hh}), minute (\texttt{mm}), and second (\texttt{ss}). \texttt{<T>} is the temperature in unit K. The \ExoMolHR database calculates the intensity at specified temperatures and is filtered by wavenumber range and minimal intensity. The downloaded data is provided in \texttt{CSV} format and uses the column names given in Table \ref{tab:csvfmt} as the header.

\begin{deluxetable}{lllll}
\tabletypesize{\small}
\tablecolumns{3} 
\setlength{\tabcolsep}{1.5mm}{
\tablecaption{Quantum number labels for each isotopologue in the \ExoMolHR database. \label{tab:QNJlist}}
\tablehead{
\colhead{Molecule} & \colhead{Isotopologue} & \colhead{Quantum number label} 
}
\startdata
AlCl       & $^{27}$Al$^{35}$Cl, $^{27}$Al$^{37}$Cl & $+/-$ \ e$/$f \ ElecState \ $v$ \ $\Lambda$ \ $\Sigma$ \ $\Omega$ \\           
AlH        & $^{27}$Al$^1$H     & \\
AlO        & $^{26}$Al$^{16}$O, $^{27}$Al$^{16}$O, $^{27}$Al$^{17}$O, $^{27}$Al$^{18}$O & \\
BeH        & $^9$Be$^1$H, $^9$Be$^2$H & \\
C$_2$      & $^{12}$C$_2$       & \\
CaH        & $^{40}$Ca$^1$H     & \\
CN         & $^{12}$C$^{14}$N   & \\
MgH        & $^{24}$Mg$^1$H, $^{25}$Mg$^1$H, $^{26}$Mg$^1$H & \\
NO         & $^{14}$N$^{16}$O   & \\
PN         & $^{31}$P$^{14}$N   & \\
SiN        & $^{28}$Si$^{14}$N, $^{28}$Si$^{15}$N, $^{29}$Si$^{14}$N, $^{30}$Si$^{14}$N & \\
SiO        & $^{28}$Si$^{16}$O  & \\
TiO        & $^{47}$Ti$^{16}$O  & \\
VO         & $^{51}$V$^{16}$O   & \\
YO         & $^{89}$Y$^{16}$O, $^{89}$Y$^{17}$O, $^{89}$Y$^{18}$O & \\
ZrO        & $^{90}$Zr$^{16}$O, $^{91}$Zr$^{16}$O, $^{92}$Zr$^{16}$O, $^{93}$Zr$^{16}$O, $^{94}$Zr$^{16}$O, $^{96}$Zr$^{16}$O & \\
SO         & $^{32}$S$^{16}$O     & $+/-$ \ ElecState \ $v$ \ $\Lambda$ \ $\Sigma$ \ $\Omega$ \\
LiOH       & $^6$Li$^{16}$O$^1$H, $^7$Li$^{16}$O$^1$H  & $+/-$ \ e$/$f \ ElecState \ $v_1$ \ $v_2$ \ $l_2$ \ $v_3$  \\
NH         & $^{14}$N$^1$H, $^{14}$N$^2$H, $^{15}$N$^1$H, $^{15}$N$^2$H & $+/-$ \ e$/$f \ ElecState \ $v$ \ $\Lambda$ \ N \ F1$/$F2$/$F3 \\
CaOH       & $^{40}$Ca$^{16}$O$^1$H & $+/-$ \ e$/$f \ N \ ElecState \ L \ $v_1$ \ $v_2$ \ $l_2$ \ $v_3$ \ $\Omega$ \ F1$/$F2$/$F3 \\
SO$_2$     & $^{32}$S$^{16}$O$_2$ & $+/-$ \ $K_a$ \ $K_c$ \ $v_1$ \ $v_2$ \ $v_3$ \\
H$_2$O     & $^1$H$_2$$^{16}$O & G$_{\rm{rve}}$ \ $v_1$ \ $v_2$ \ $v_3$ \ $K_a$ \ $K_c$ \\
H$_2$S     & $^1$H$_2$$^{32}$S & \\
H$_3$$^+$ & $^1$H$_3$$^+$, $^2$H$_3$$^+$ & e$/$f \ G$_{\rm{rve}}$ \ N \ Isomer \ $v_1$ \ $v_2$ \ $l_2$ \ G \ U \ K \\
& $^1$H$_2$$^2$H$^+$, $^2$H$_2$$^1$H$^+$ & $+/-$ \ G$_{\rm{rve}}$ \ N \ Isomer \ $v_1$ \ $v_2$ \ $v_3$ \ $K_a$ \ $K_c$ \\
NH$_3$     & $^{14}$N$^1$H$_3$ & $+/-$ \ G$_{\rm{tot}}$ \ N \ $n_1$ \ $n_2$ \ $n_3$ \ $n_4$ \ $l_3$ \ $l_4$ \ $\tau_i$ \ K \ G$_{\rm{rot}}$ \\
& $^{15}$N$^1$H$_3$ & G$_{\rm{tot}}$ \ $n_1$ \ $n_2$ \ $n_3$ \ $l_3$ \ $n_4$ \ $l_4$ \ G$_{\rm{vib}}$ \ K \ G$_{\rm{rot}}$ \\
C$_2$H$_2$ & $^{12}$C$_2$$^1$H$_2$ & G$_{\rm{tot}}$ \ K \ e$/$f \ G$_{\rm{rot}}$ \\
N$_2$O     & $^{14}$N$_2$$^{16}$O & G$_{\rm{tot}}$ \ $n_1$ \ $n_2lin$ \ $l_2$ \ $n_3$ \ P \ N \\
H$_3$O$^+$ & $^1$H$_3$$^{16}$O$^+$ & G$_{\rm{tot}}$ \ $n_1$ \ $n_2$ \ $n_3$ \ $l_3$ \ $n_4$ \ $l_4$ \ K \ G$_{\rm{rot}}$ \\
CH$_4$     & $^{12}$C$^1$H$_4$ & G$_{\rm{tot}}$ \ $n_1$ \ $n_2$ \ $l_2$ \ $n_3$ \ $l_3$ \ M$_3$ \ $n_4$ \ $l_4$ \ M$_4$ \ n \\ 
CO$_2$     & $^{12}$C$^{16}$O$_2$ & G$_{\rm{tot}}$ \ e$/$f \ $n_1$ \ $n_2$ \ $l_2$ \ $n_3$ \ $m_1$ \ $m_2$ \ $m_3$ \ $m_4$ \ $m_5$ \\
H$_2$CO    & $^1$H$_2$$^{12}$C$^{16}$O & G$_{\rm{tot}}$ \ $v_1$ \ $v_2$ \ $v_3$ \ $v_4$ \ $v_5$ \ $v_6$ \ $K_a$ \ $K_c$ \\
H$_2$CS    & $^1$H$_2$$^{12}$C$^{32}$S & G$_{\rm{tot}}$ \ $v_1$ \ $v_2$ \ $v_3$ \ $v_4$ \ $v_5$ \ $v_6$ \ $K_a$ \ $K_c$ \\
OCS        & $^{16}$O$^{12}$C$^{32}$S & G$_{\rm{tot}}$ \ $v_1$ \ $v_2$ \ $l_2$ \ $v_3$ \ e$/$f \ G$_{\rm{vib}}$ \ G$_{\rm{rot}}$ \ Coef \ $n_1$ \ $n_2$ \ $n_3$ \\
\enddata
}
\tablecomments{Electronic state term values are defined using \textsc{PyValem} format \citep{pyvalem,pyvalem2}.
}
\end{deluxetable}


\begin{deluxetable}{llll}
\tabletypesize{\small}
\tablecolumns{4} 
\setlength{\tabcolsep}{2.4mm}{
\tablecaption{The formats and descriptions corresponding to quantum number labels. \label{tab:QNfmt}}
\tablehead{
\colhead{Label \ \ \ \ \ \ \ \ \ \ \ \ \ \ \ } & \colhead{Fortran Format} & \colhead{C Format} & \colhead{Description} 
}
\startdata
$J$ & I7/F7.1 & \%7d/\%7.1f & Total angular momentum \\
& I7 & \%7d & AlCl, AlH, C$_2$, C$_2$H$_2$, CH$_4$, CO$_2$, H$_2$CO, H$_2$CS, H$_2$O, H$_2$S, H$_3$$^+$ \\
& & & H$_3$O$^+$, LiOH, N$_2$O, NH, NH$_3$, OCS, PN, SiO, SO, SO$_2$, TiO, VO \\
& F7.1 & \%7.1f & AlO, BeH, CN, CaH, CaOH, MgH, NO, SiN, YO, ZrO \\
$+/-$ & A1 & \%1s & Total parity \\
e$/$f & A1 & \%1s & Rotationless parity \\
ElecState & A12 &\%12s & Electronic term value \\
$v$ & I3 & \%3d & Vibrational quantum number \\
$|\Lambda|$ & I3 & \%3d & Absolute value of the projection of electronic angular momentum \\
$|\Sigma|$ & I3 & \%3d & Absolute value of the projection of the electronic spin \\
$|\Omega|$ & I5/F5.1 & \%5d/\%5.1f & Absolute value of the projection of the total angular momentum \\
& I5 & \%5d & AlCl, AlH, C$_2$, TiO, ZrO \\
& F5.1 & \%5.1f & AlO, BeH, CN, CaH, CaOH, MgH, NO, PN, SO, SiN, SiO, VO, YO \\
G$_{\rm{tot}}$ & A3 & \%3s & Total symmetry \\
G$_{\rm{rot}}$ & A3 & \%3s & Symmetry of rotational contribution \\
G$_{\rm{rve}}$ & A3 & \%3s & Rovibrational symmetry label for upper and lower state\\
G$_{\rm{vib}}$ & A3 & \%3s & Symmetry of vibrational contribution \\
N & I5 & \%5d & Vibrational state ID or rotational angular momentum\\
L & I3 & \%3d & Vibronic angular momentum quantum number \\
P & I2 & \%2d & Polyad number \\
F & I2 & \%2d & Fine structure counting number \\
F1$/$F2$/$F3 & I2 & \%2d & Spin components F1 and F3 \\
U & A1 & \%1s & U-notation of \citet{84Watson.H3+} \\
G & I2 & \%2d & Absolute value of $k-l_2$ \\
K & I2 & \%2d & State projection of the \citet{84Watson.H3+} rotational quantum number \\
$K_a$ & I2 & \%2d & $K_a$ rotational quantum number \\
$K_c$ & I2 & \%2d & $K_c$ rotational quantum number \\
$M_3$, $M_4$ & I2 & \%2d & Multiplicity index quantum numbers \\
$v_1$,$v_2$,$v_3$,$v_4$,$v_5$,$v_6$ & I2 & \%2d & $v_i$ Local mode vibrational quantum numbers \\
$l_2$,$l_3$,$l_4$ & I2 & \%2d & Normal mode vibrational angular momentum quantum number \\
$n_1$,$n_2$,$n_3$,$n_4$ & I2 & \%2d & Normal mode vibrational quantum numbers \\
$m_1$,$m_2$,$m_3$,$m_4$,$m_5$ & I2 & \%2d & HITRAN normal mode quantum number \cite{jt857}\\
$n$ & I2 & \%2d & Polyad or rotational counting number \\
$n_2lin$ & I2 & \%2d & Normal mode linear molecule bending quantum number\\
$\tau_i$ & I1 & \%1d & Inversion parity (0 or 1) \\
Isomer & A1 & \%1s & Nuclear spin isomer \\
Coef & F5.2 & \%5.2f & Coefficient with the largest contribution to the ($J=0$) contracted set \\
\enddata
}
\tablecomments{Electronic state term values are defined using \textsc{PyValem} format \citep{pyvalem,pyvalem2}. The descriptions of the labels in the table are for reference only. Sometimes there may be slight variations in their usage. More accurate and detailed descriptions are available in the definition file, \texttt{.def.json}, for  the recommended dataset for each isotopologue. The meaning of each label can be further understood by referring to the notes under the example tables in the corresponding line list papers.
}
\end{deluxetable}

Table~\ref{tab:QNJlist} displays the quantum number labels and the format of the quantum number $J$ for each isotopologue in the \ExoMolHR database. Table~\ref{tab:QNfmt} details the formats and descriptions of quantum numbers for various quantum number labels. The electronic state term values in both \textsc{ExoMol} and \ExoMolHR databases are defined using \textsc{PyValem} format. 
\textsc{PyValem} \citep{pyvalem,pyvalem2} is a Python package designed for parsing, validating, manipulating, and interpreting chemical formulas, quantum states, and labels of atoms, ions, and small molecules. For example, X$^1\,\Sigma^+$ is written as \texttt{X(1SIGMA+)}.
Sometimes, due to conventional differences in the notation used for the quantum number labels of different molecules in the \textsc{ExoMol} definition files (\texttt{.def}), there differences in the notation  adopted between molecules meaning that  $v_i$, $n_i$ or $m_i$ can represent similar quantum numbers.

\subsection{Use cases} \label{sec:usecases}

\begin{figure}
\centering
\gridline{\fig{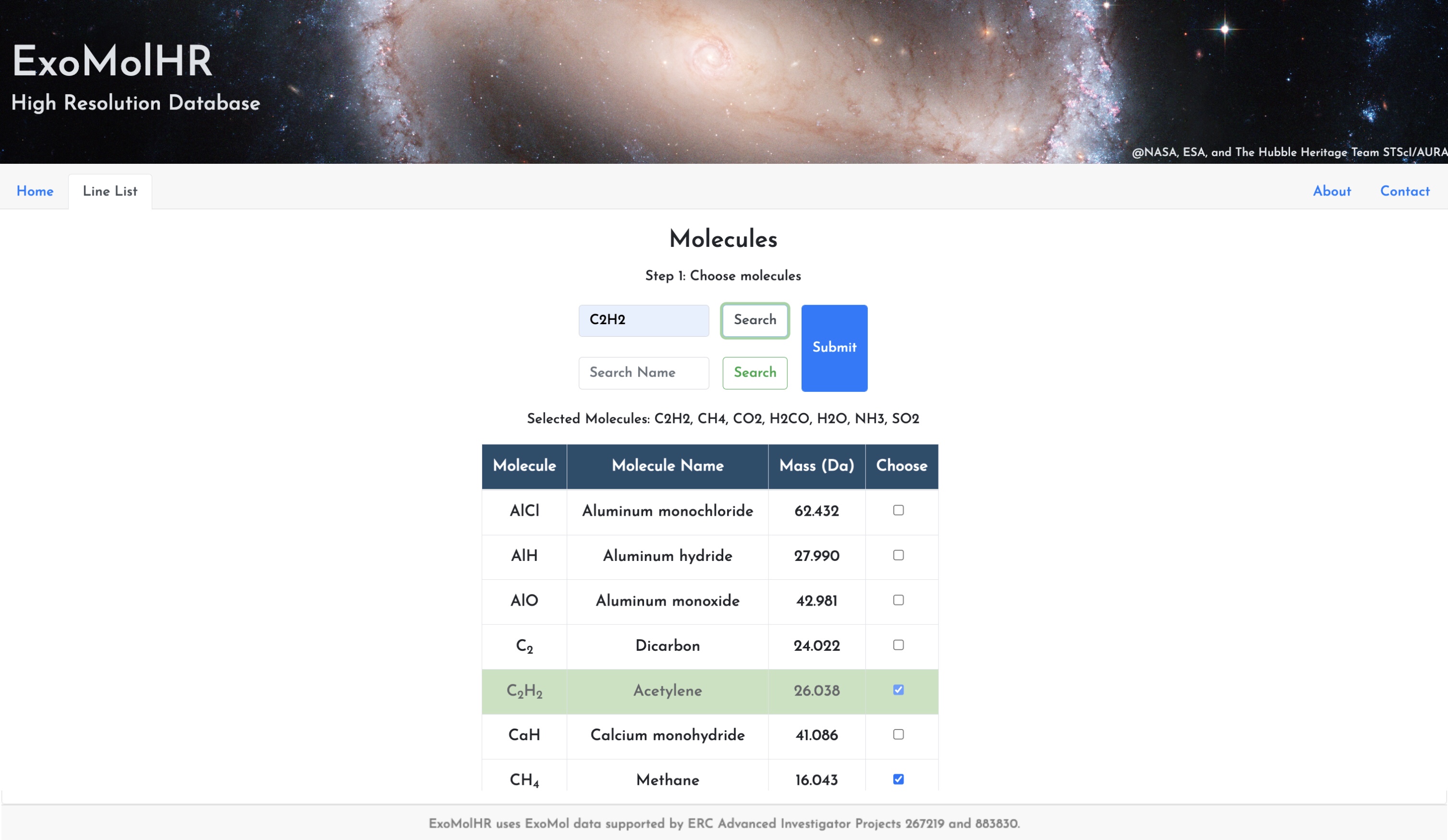}{0.5\textwidth}{Step 1. Choose molecules}
          \fig{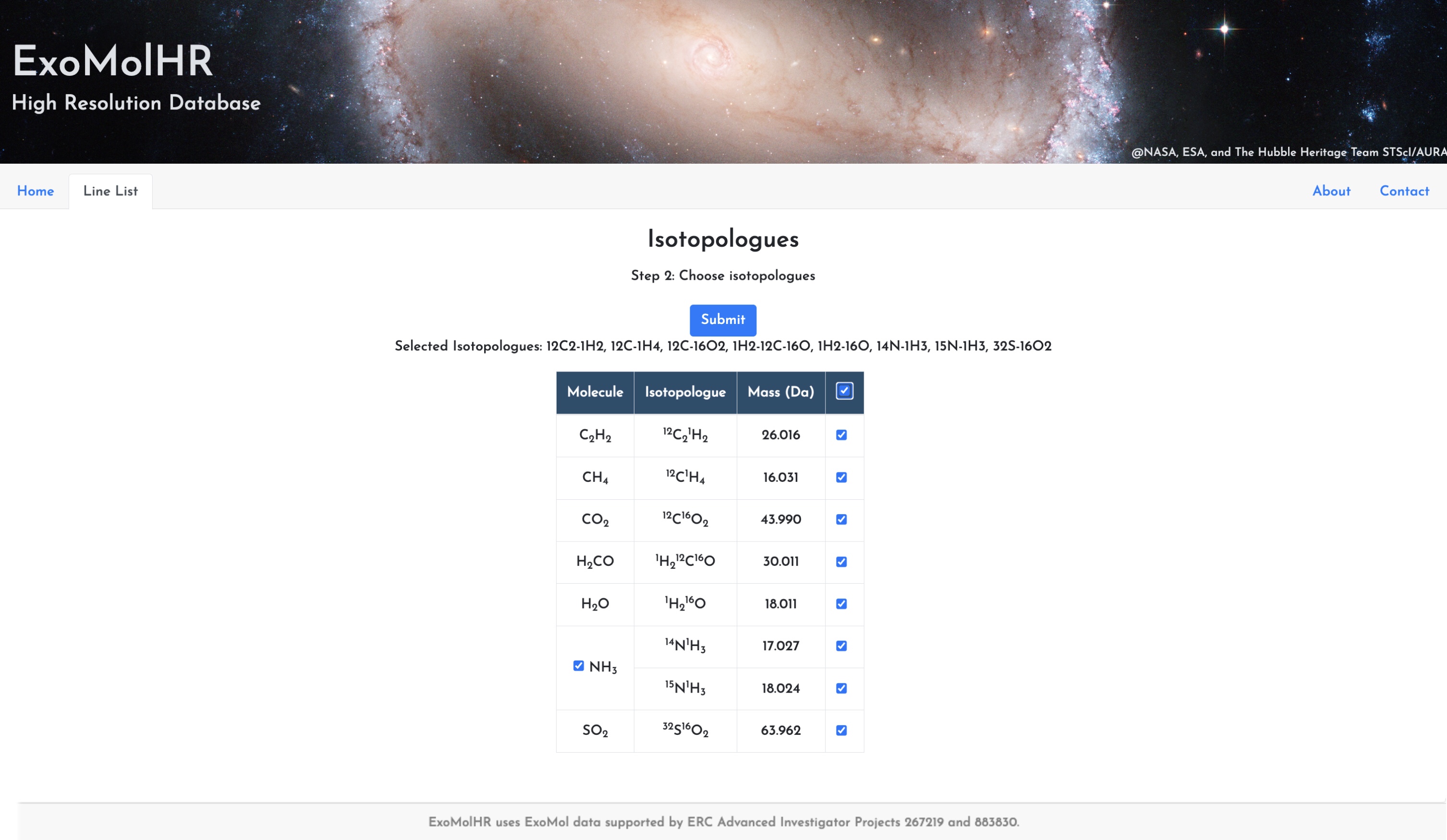}{0.5\textwidth}{Step 2. Choose isotopologues}}
\gridline{\fig{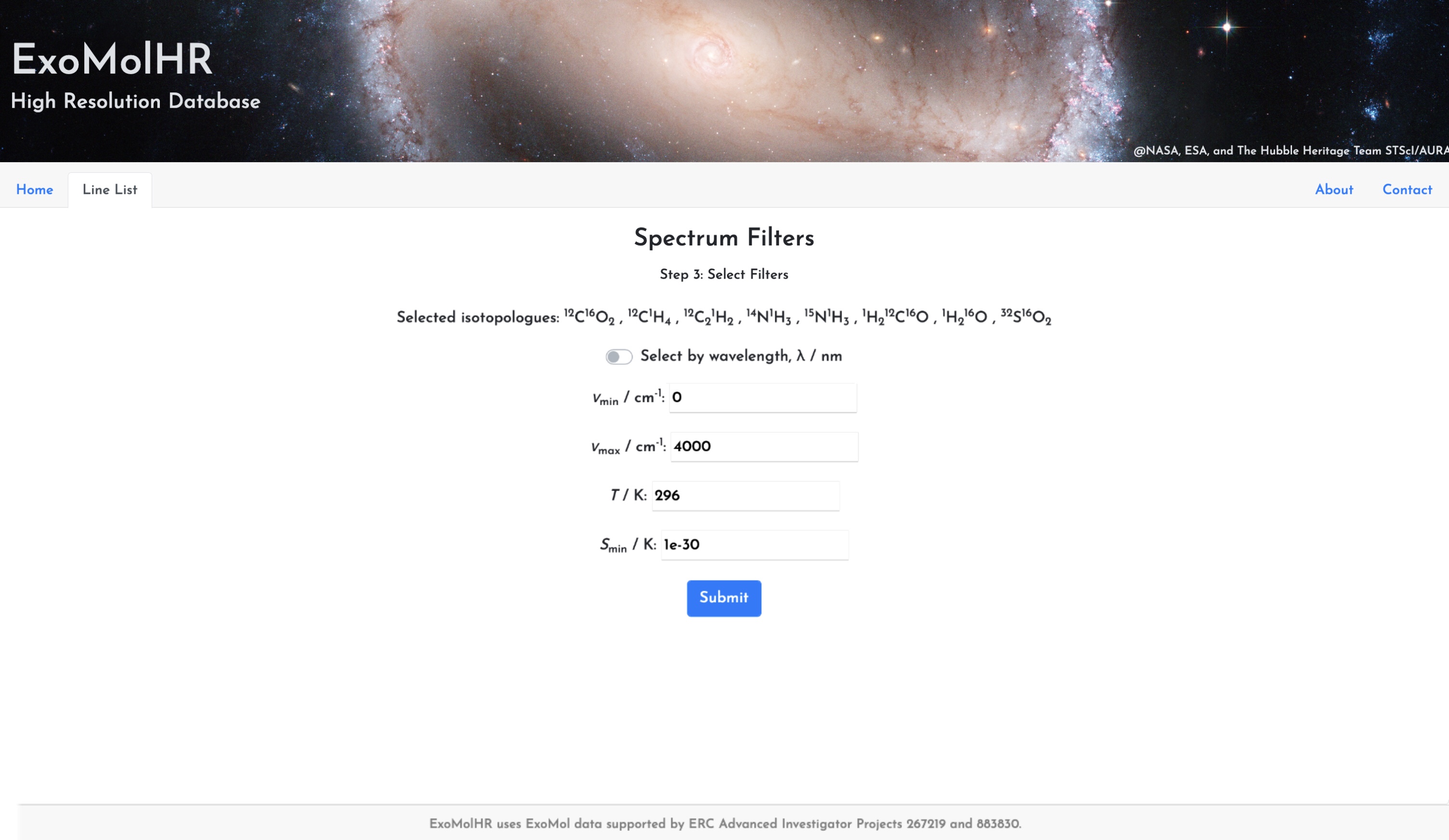}{0.5\textwidth}{Step 3. Select filters}
          \fig{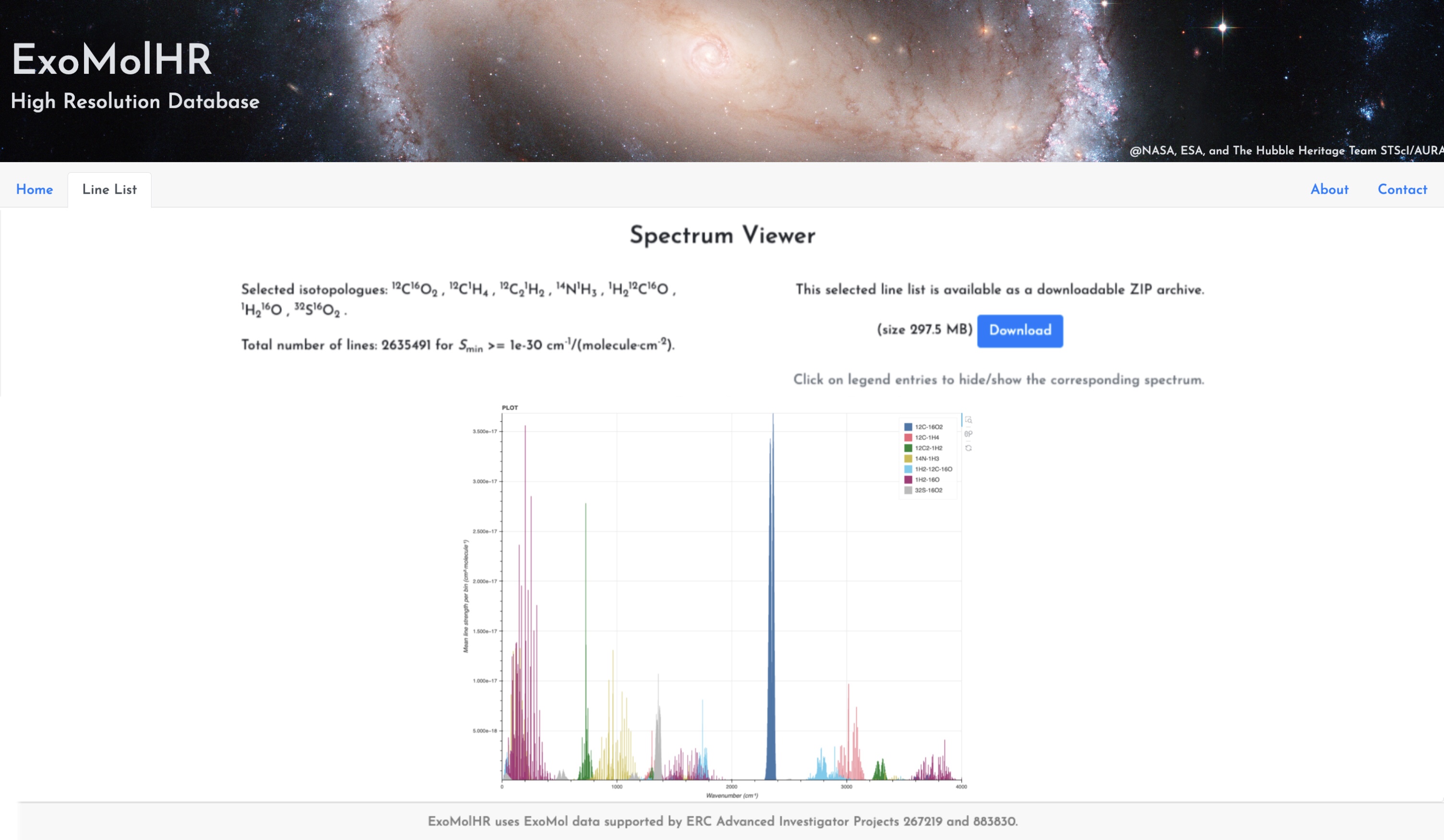}{0.5\textwidth}{Step 4. Spectrum viewer}}
\caption{Four steps for getting the intensity file from the \ExoMolHR website. \label{fig:HRWebStep}}
\end{figure}

\ExoMolHR website provides a download link to the \texttt{.csv} file(s) which contain the calculated results,  and plots the stick spectra for one or multiple isotopologues. Users can use the following steps to obtain the final results. Figure~\ref{fig:HRWebStep} shows the screenshots of these four steps.
\begin{enumerate}[leftmargin=1.55cm, label=Step \arabic*.]
\item Molecules: Users can choose one or multiple molecules from the form. There are 2 search boxes for searching by molecule name and formula. 
\item Isotopologues: Similarly, one or more isotopologues can be selected in this step; users can select all isotopologues by clicking the "select all" button.
\item Spectral filters: Users can define the wavenumber range by the given $\nu_{\rm{min}}$ and $\nu_{\rm{max}}$ in cm$^{-1}$ or wavelength range by $\lambda_{\rm{min}}$ and $\lambda_{\rm{max}}$ in $\mu$m, temperature $T$ in K, and minimum intensity $S_{\rm{min}}$ in cm$/$molecule. 
\item Spectrum viewer: finally, users are provided with a plot of intensities calculated at the given temperature for one or multiple isotopologues in different colours for the requested wavenumber range. The webpage also provides the number of lines and file size for total result files. Users can download a compressed \texttt{.zip} file and the result files are named separately for each species/isotopologues. The structure of this file is specified in Section~\ref{sec:filestructure}.
\end{enumerate}

Note the \ExoMolHR website only plots stick spectra. The \textsc{ExoMol} database provides a set of pressure broadening
parameters \citep{jt684,jt919,jt957}. These parameters can be combined with the \texttt{.csv} file to provide temperature and pressure-dependent cross sections. The python program \textsc{PyExoCross} \citep{jt914} is being adapted ingest the intensities from the \ExoMolHR database and the broadening parameters from the \textsc{ExoMol} database to calculate the cross sections. However, we note that for most problems this process should probably
use the original  \textsc{ExoMol} line list to ensure the inclusion of all lines in the spectral region of interest.

\subsection{Application programming interface (API)} \label{sec:API}

Users can use the application programming interface (API) to jump to the spectrum viewer website. The API format is \texttt{https://www.exomol.com/exomolhr/get-data/?numin=<$\nu_{\rm{min}}$>\&numax=<$\nu_{\rm{max}}$>\&T=<$T$>\&Smin=<$S_{\rm{min}}$>\&iso=<ISO-SL\\UG$_{\rm{i}}$>}. If multiple isotopologues are requested, then  \texttt{\&iso=<ISO-SLUG$_{\rm{i}}$>} should be used repeatedly to specify these different isotopologues. For example,  \texttt{\&iso=<ISO-SLUG$_{\rm{1}}$>\&iso=<ISO-SLUG$_{\rm{2}}$>}. As an example, 
a user request for \ExoMolHR files with intensities calculated at temperature $T=296$ K with an intensity  threshold for $S_{\rm{min}}=10^{-30}$ cm$/$molecule for $^1$H$_2$$^{16}$O, $^{12}$C$^{16}$O$_2$, and $^{32}$S$^{16}$O$_2$ in the wavenumber range from $\nu_{\rm{min}}=0$ to $\nu_{\rm{max}}=1000$ cm$^{-1}$, the API is \href{https://www.exomol.com/exomolhr/get-data/?numin=0&numax=1000&T=296&Smin=1e-30&iso=1H2-16O&iso=12C-1H4&iso=32S-16O2}{https://www.exomol.com/exomolhr/get-data/?numin=0\&numax=1000\&T=296\&Smin=1e-30\&iso=1H2-16O\&iso=12C-1H4\&iso=32S-16O2}. Note that in the file name the species are ordered in the order of the original request.

The \ExoMolHR compressed \texttt{.zip} file is named with the creation date and time which has 14 digits in format yyyymmddhhmmss (year, month, day, hour, minute, and second), for example, \href{https://www.exomol.com/exomolhr/get-data/download/?archive\_name=20240907222904.zip}{https://www.exomol.com/exomolhr/get-data/download/?archive\_name=20240907222904.zip}. 
This means users cannot directly download files via the API, but it can still be downloaded by clicking the download button on the spectrum viewer website. This method is typically more intuitive and ensures that users can see the data visualization and file details they are about to download before doing so, which helps reduce the likelihood of erroneous downloads.

\subsection{Database summary and sources} \label{sec:dbsum}

Table~\ref{tab:HRlist} displays a list of molecules, isotopologues, datasets, and their corresponding versions (update date) and data sources. The number of initial states in the \textsc{ExoMol} database \texttt{.states} file as a data source and the number of the states with lower uncertainty ($\Delta E \le 0.01$ cm$^{-1}$) used in the \ExoMolHR database are compared in this table. Similarly,  Table~\ref{tab:HRlist} lists the number of transitions contained the \texttt{.trans} file in the \textsc{ExoMol} database and the total number of transitions and file size for each dataset compared to the final line list number and file size in the \ExoMolHR database.

\section{Comparisons and discussion} \label{sec:results}

Fifteen of the \nmol\ species currently considered by \ExoMolHR are also present in the HITRAN database \citep{jt857}.
To compare \ExoMolHR results with its initial source database \textsc{ExoMol} and high-resolution database HITRAN, we present the following figures in this section calculated by a Python program \PyExoCross \citep{jt914}. 
To mimic HITRAN, we only consider those lines whose intensities are larger than $10^{-30}$ cm$/$molecule at $T=296$ K.
As can be seen, the coverage provided by \ExoMolHR is generally more complete than that provided by HITRAN.

\begin{figure}[ht!]
\centering
\gridline{\fig{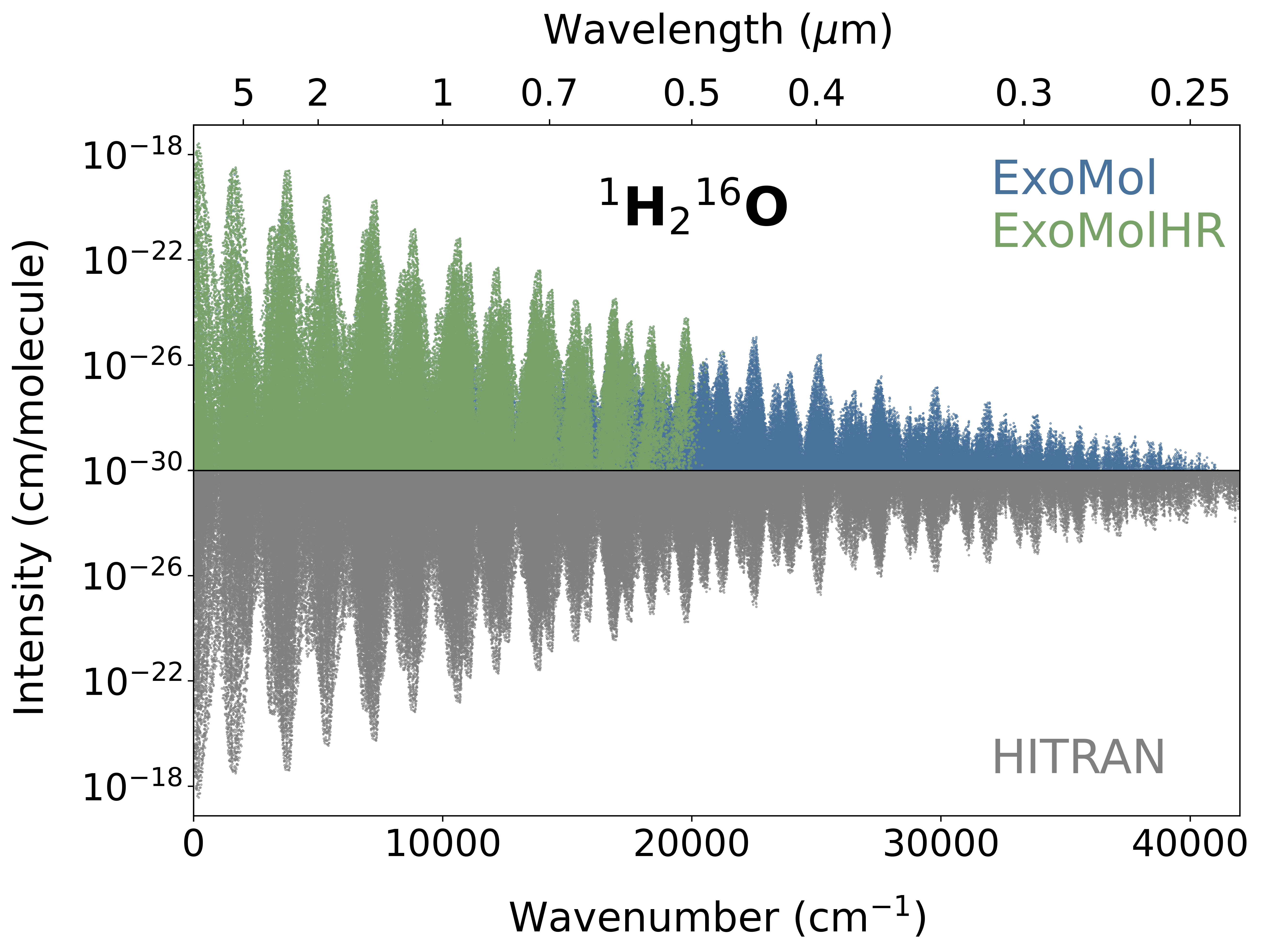}{0.5\textwidth}{}
          \fig{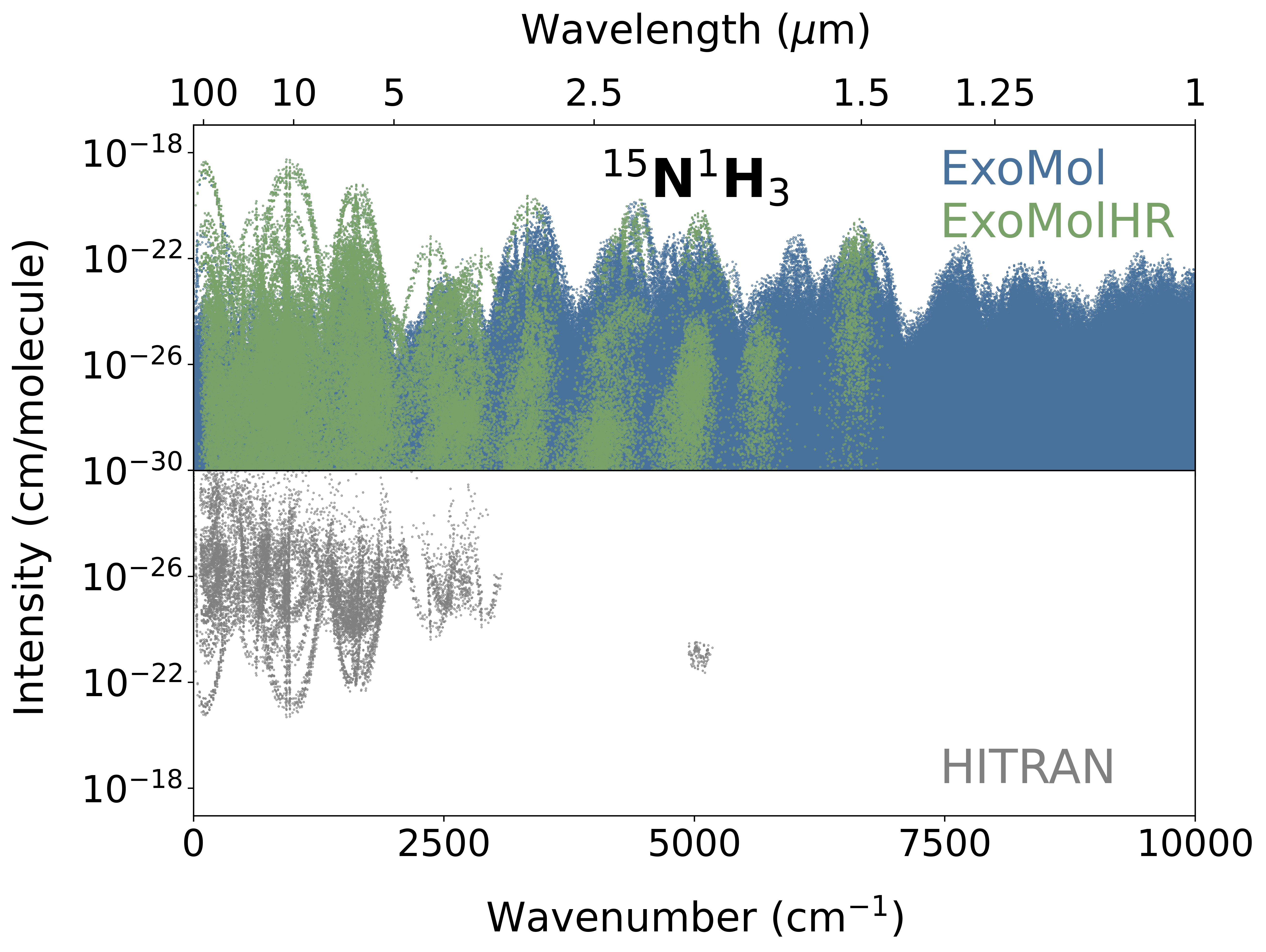}{0.5\textwidth}{}}
\gridline{\fig{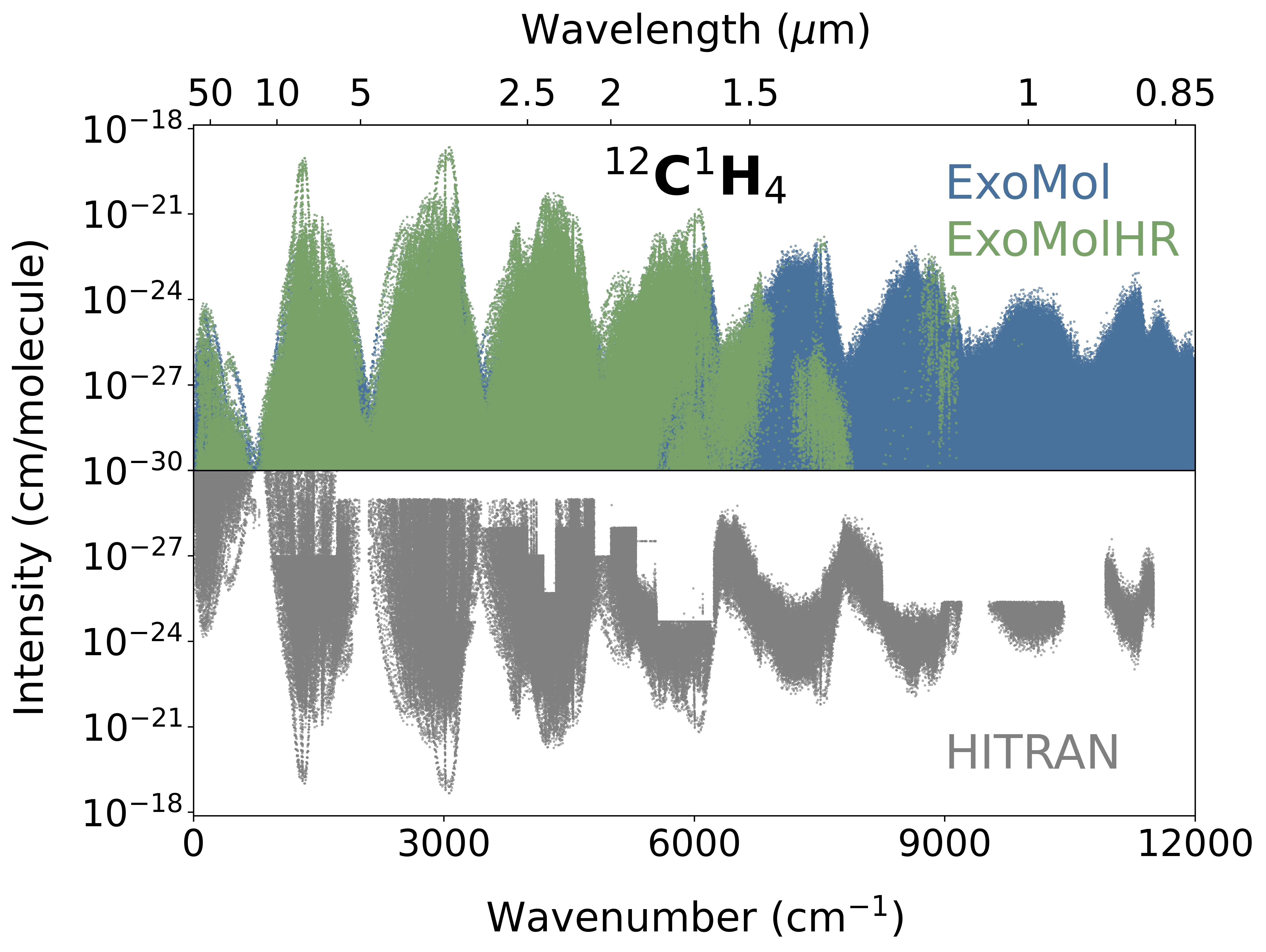}{0.5\textwidth}{}
          \fig{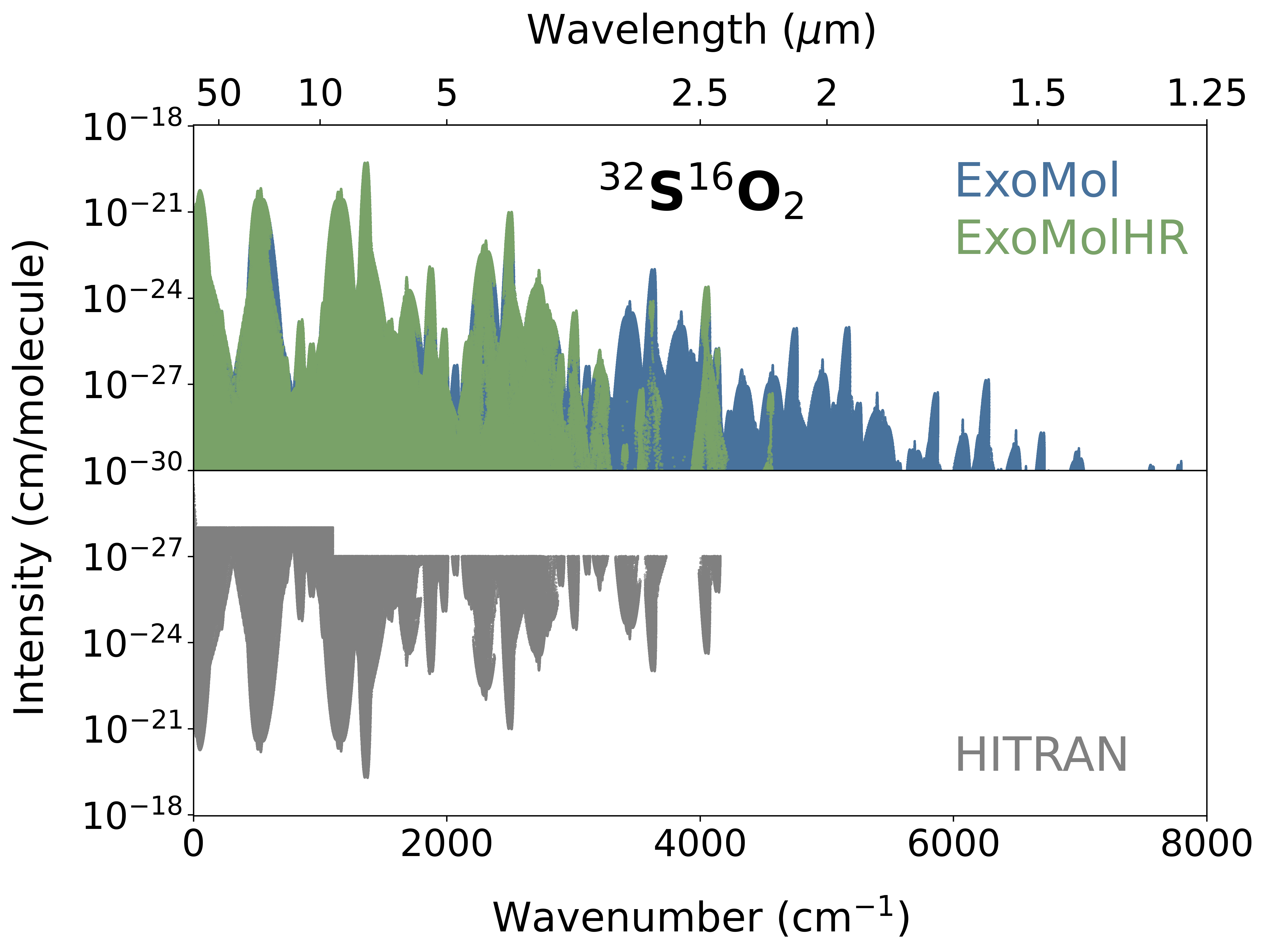}{0.5\textwidth}{}}
\caption{Comparison of the line intensities (cm$/$molecule) from ExoMol, \ExoMolHR (top) and HITRAN (bottom) databases at $T=296$ K of $^1$H$_2$$^{16}$O, $^{15}$N$^1$H$_3$, $^{12}$C$^1$H$_4$, and $^{32}$S$^{16}$O$_2$. The highlighted points in the upper panels show  intensities provided by \textsc{ExoMolHR}, while the darker background shows the full coverage given by the POKAZATEL, CoYuTe-15, MM, and ExoAmes line lists of $^1$H$_2$$^{16}$O, $^{15}$N$^1$H$_3$, $^{12}$C$^1$H$_4$, and $^{32}$S$^{16}$O$_2$, respectively. \label{fig:3dbcompare}}
\end{figure}

Figure~\ref{fig:3dbcompare} compares the intensities of H$_2$O, NH$_3$, CH$_4$, and SO$_2$ obtained from the 2024 release of \textsc{ExoMol} \citep{jt939}, \textsc{ExoMolHR}, and 2020 release of HITRAN \citep{jt857} databases. These intensities (cm$/$molecule) are calculated at a temperature of 296 K, covering the entire wavenumber range in cm$^{-1}$. 
The line list from the \ExoMolHR and \textsc{ExoMol} databases are POKAZATEL of $^1$H$_2$$^{16}$O \citep{jt734}, CoYuTe-15 of $^{15}$N$^1$H$_3$ \citep{jt952}, MM of $^{12}$C$^1$H$_4$ \citep{jt927}, and ExoAmes of $^{32}$S$^{16}$O$_2$ \citep{jt635}. 
By comparison of the distribution of the \ExoMolHR and HITRAN points with the \textsc{ExoMol} points in this figure, it is evident that high-resolution spectra are generally distributed in the low-wavenumber range at room temperature. When comparing the distribution of the \ExoMolHR\ and HITRAN  points, we observe that high-resolution spectra with low uncertainty are primarily concentrated in the low-wavenumber region.

\begin{figure}[ht!]
\centering
\gridline{\fig{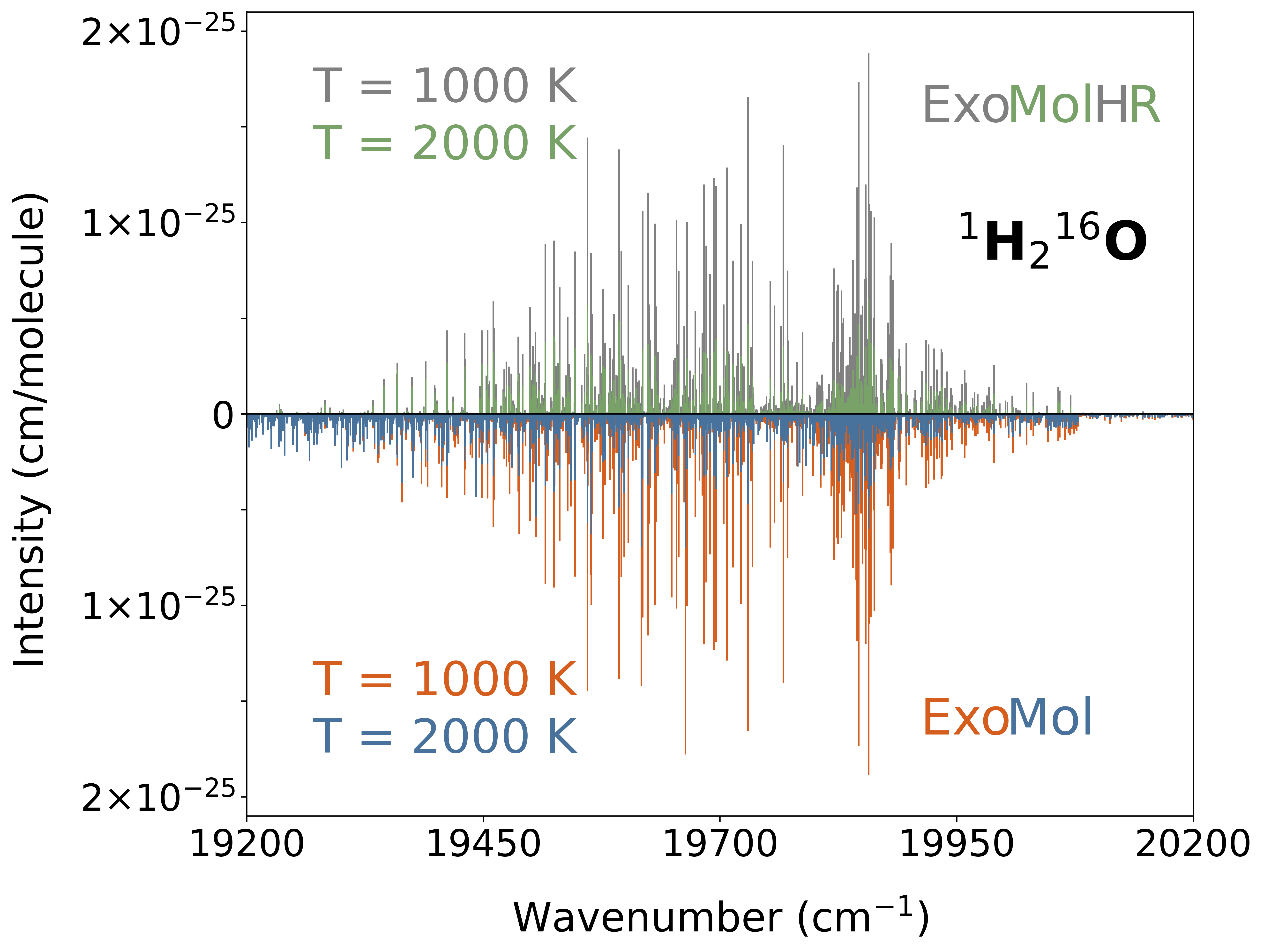}{0.5\textwidth}{}
          \fig{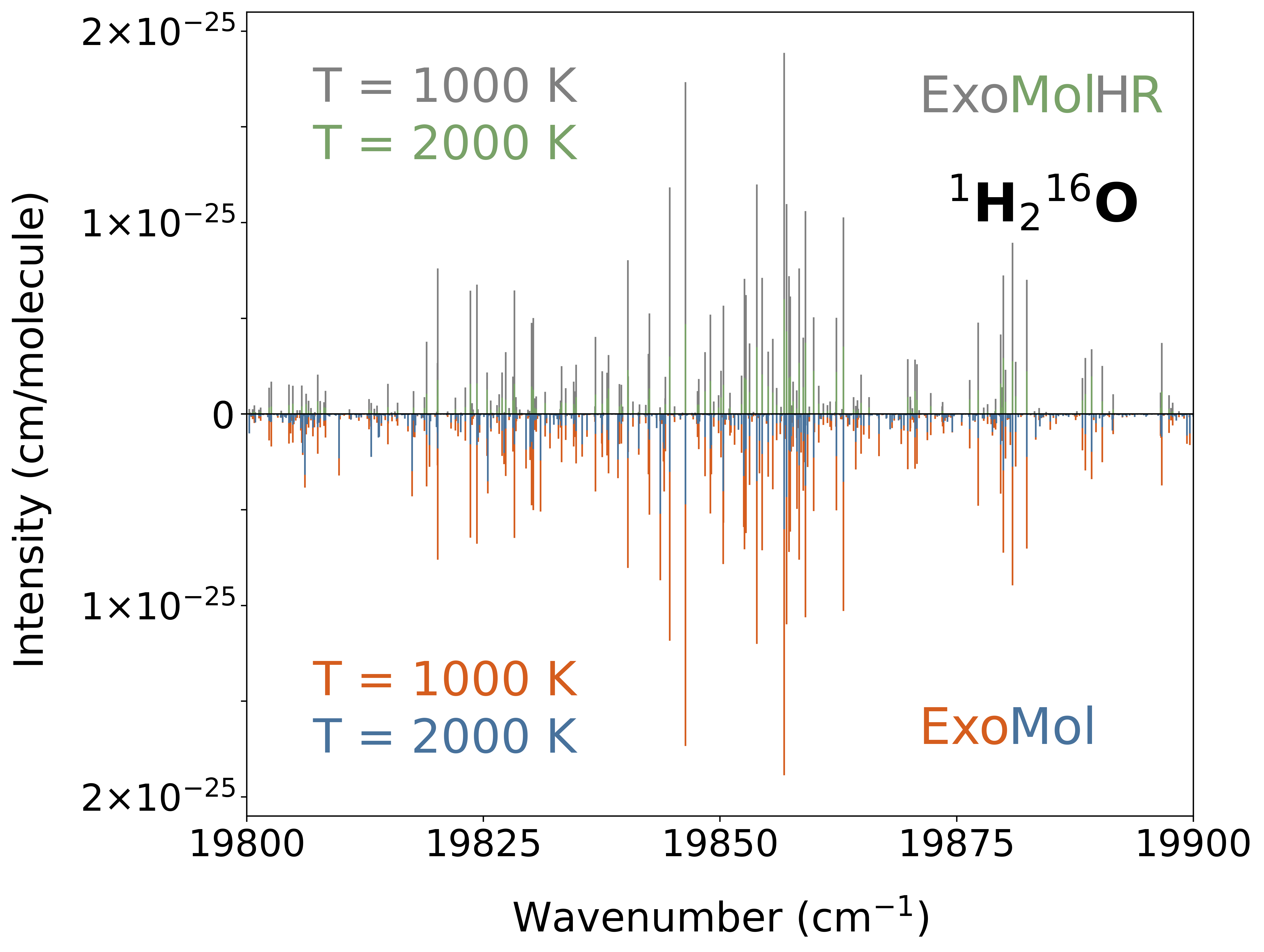}{0.5\textwidth}{}}
\gridline{\fig{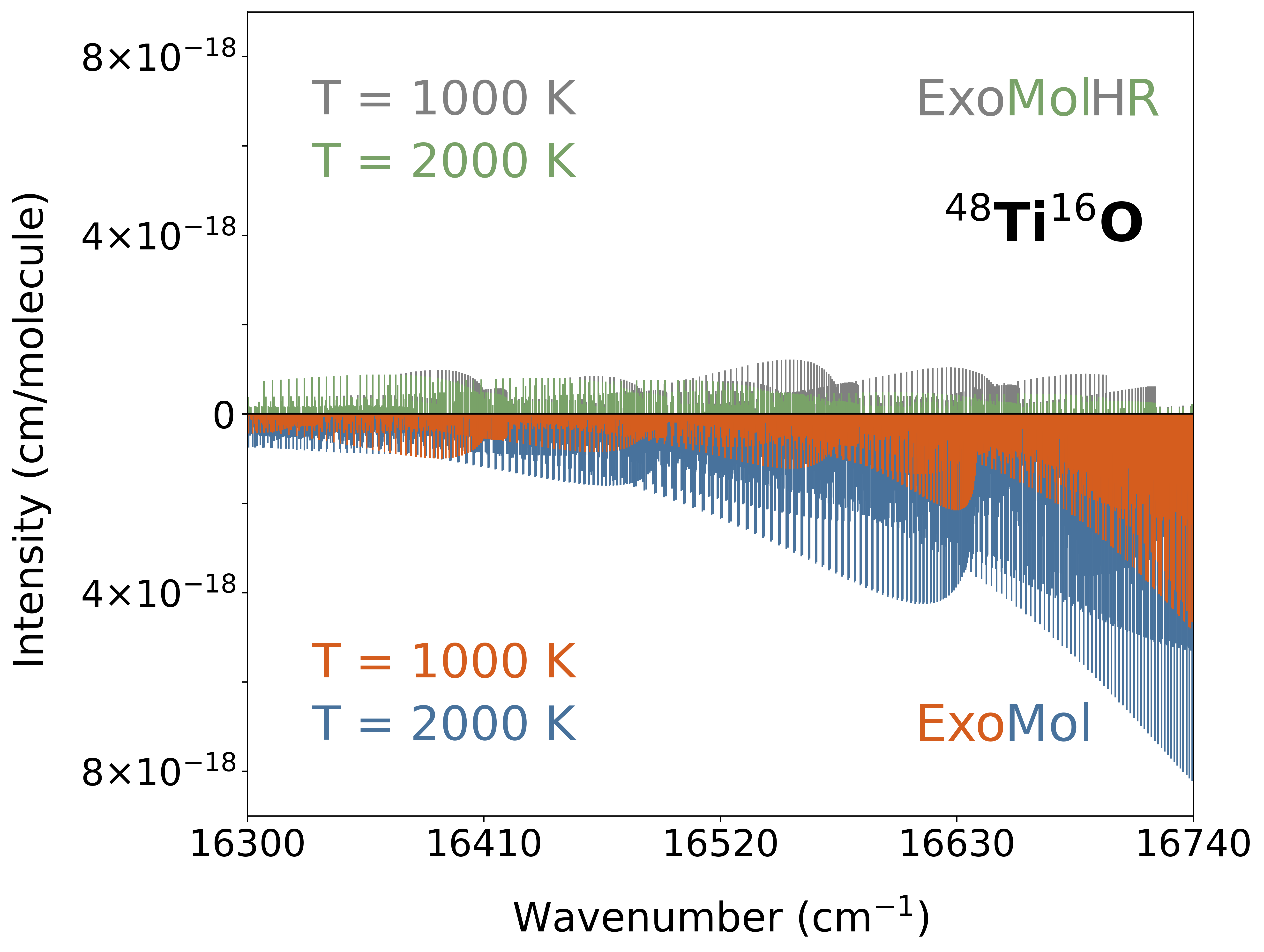}{0.5\textwidth}{}
          \fig{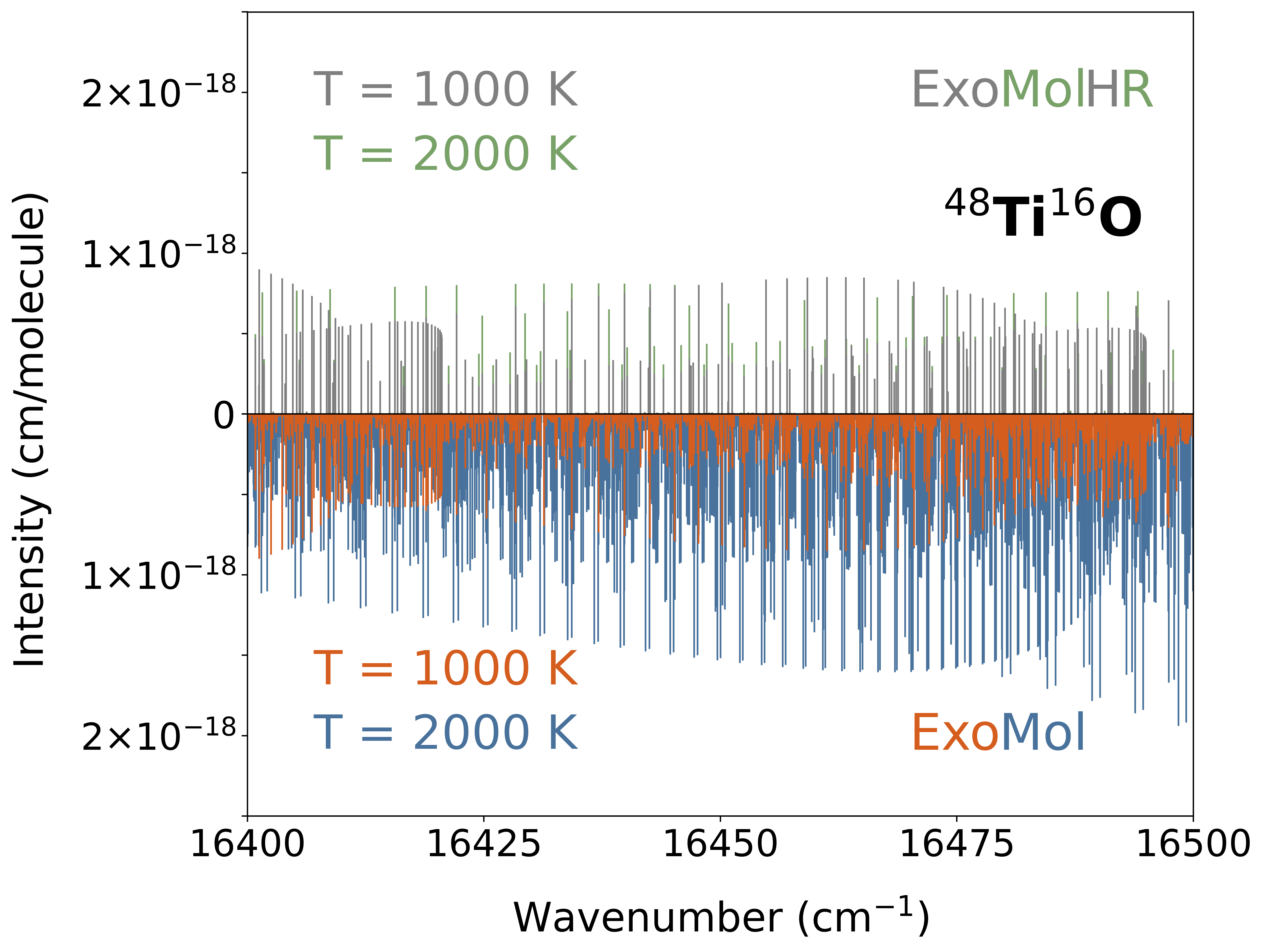}{0.5\textwidth}{}}
\gridline{\fig{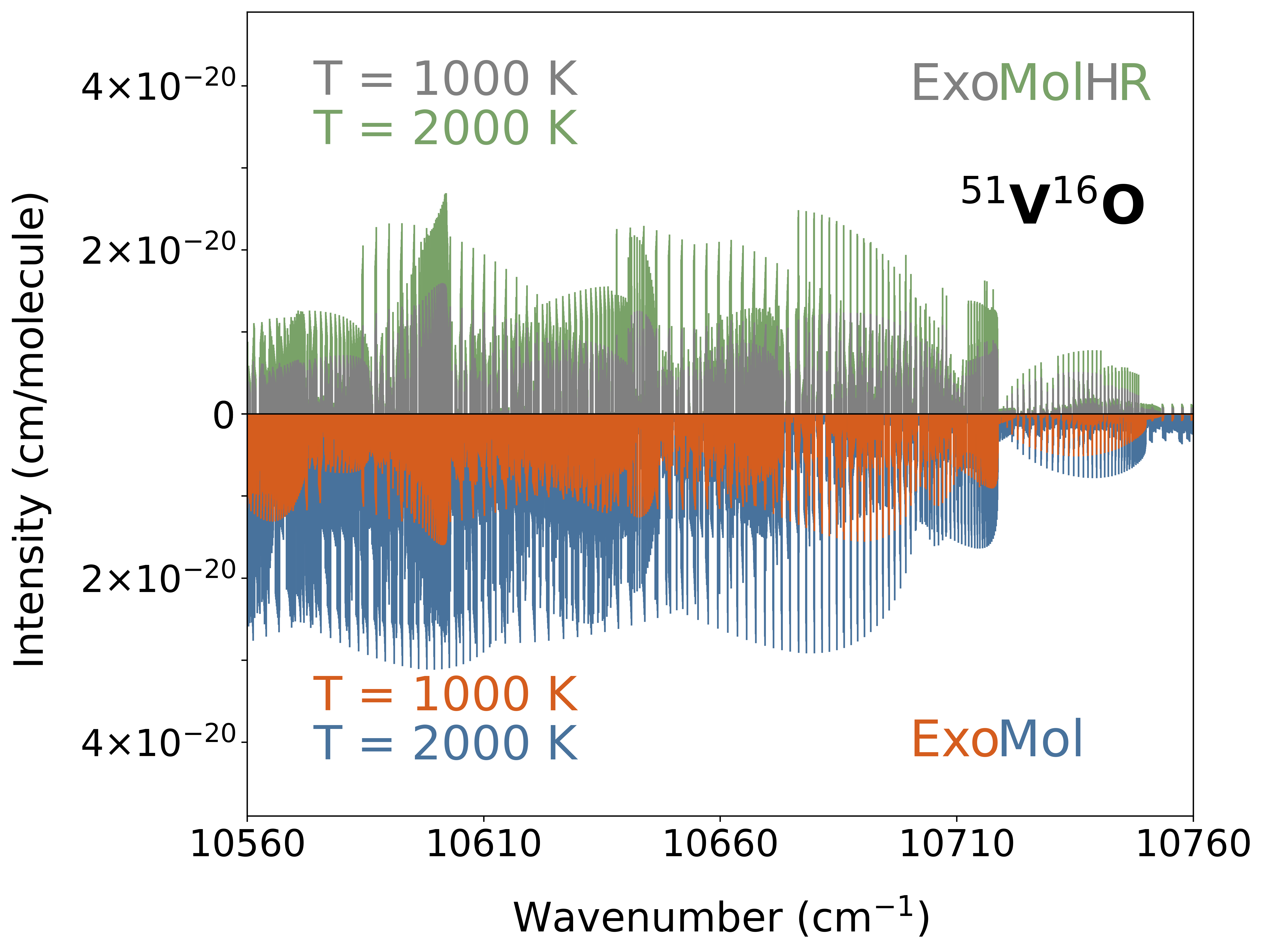}{0.5\textwidth}{}
          \fig{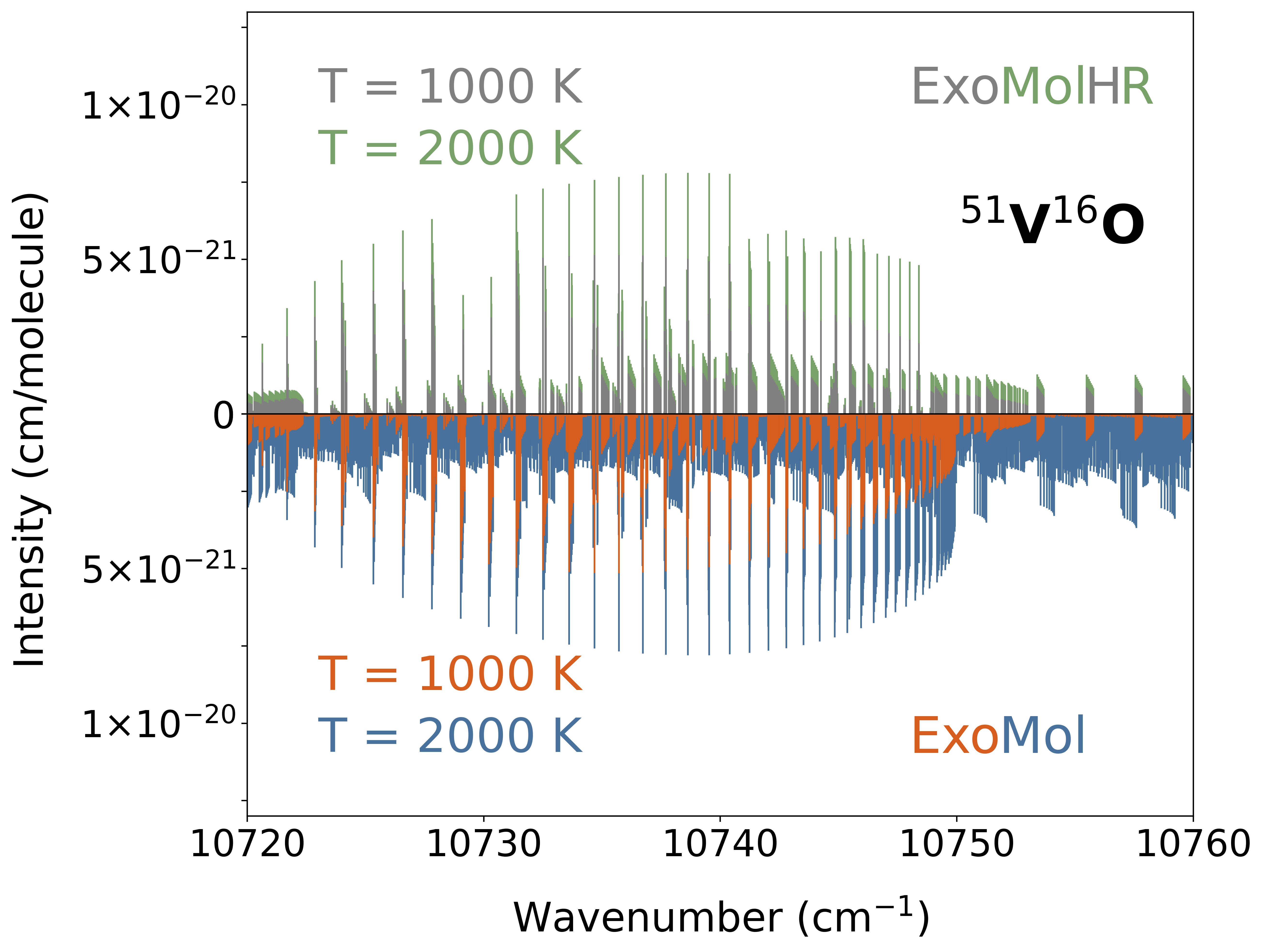}{0.5\textwidth}{}}
\caption{Comparison of the line intensities (cm$/$molecule) from \ExoMolHR (top) and \textsc{ExoMol} (bottom) databases at temperatures $T=1000$ and $2000$ K of $^1$H$_2$$^{16}$O, $^{48}$Ti$^{16}$O, and $^{51}$V$^{16}$O. \label{fig:2dbcompare}}
\end{figure}

Figure~\ref{fig:2dbcompare} presents a comparative analysis of the intensities (cm$/$molecule) for H$_2$O \citep{jt734} derived from the \textsc{ExoMol} and \ExoMolHR databases at two distinct temperatures, $T=1000$ and $2000$ K focused on the wavenumber range of $[19200,20200]$ cm$^{-1}$ and provides a detailed examination within the narrower interval of $[19800,19900]$ cm$^{-1}$. Similar comparisons  are given for $^{48}$Ti$^{16}$O \citep{jt760,jt948} and $^{51}$V$^{16}$O \citep{jt923,jt948}; these species are not considered in the HITRAN database. It can be seen that for TiO and VO, the \ExoMolHR coverage is patchier than for the better studied HITRAN molecules. Better coverage of the high resolution lines for these species will require new high resolution
spectroscopic measurements.

\begin{figure}[ht!]
\centering
\gridline{\fig{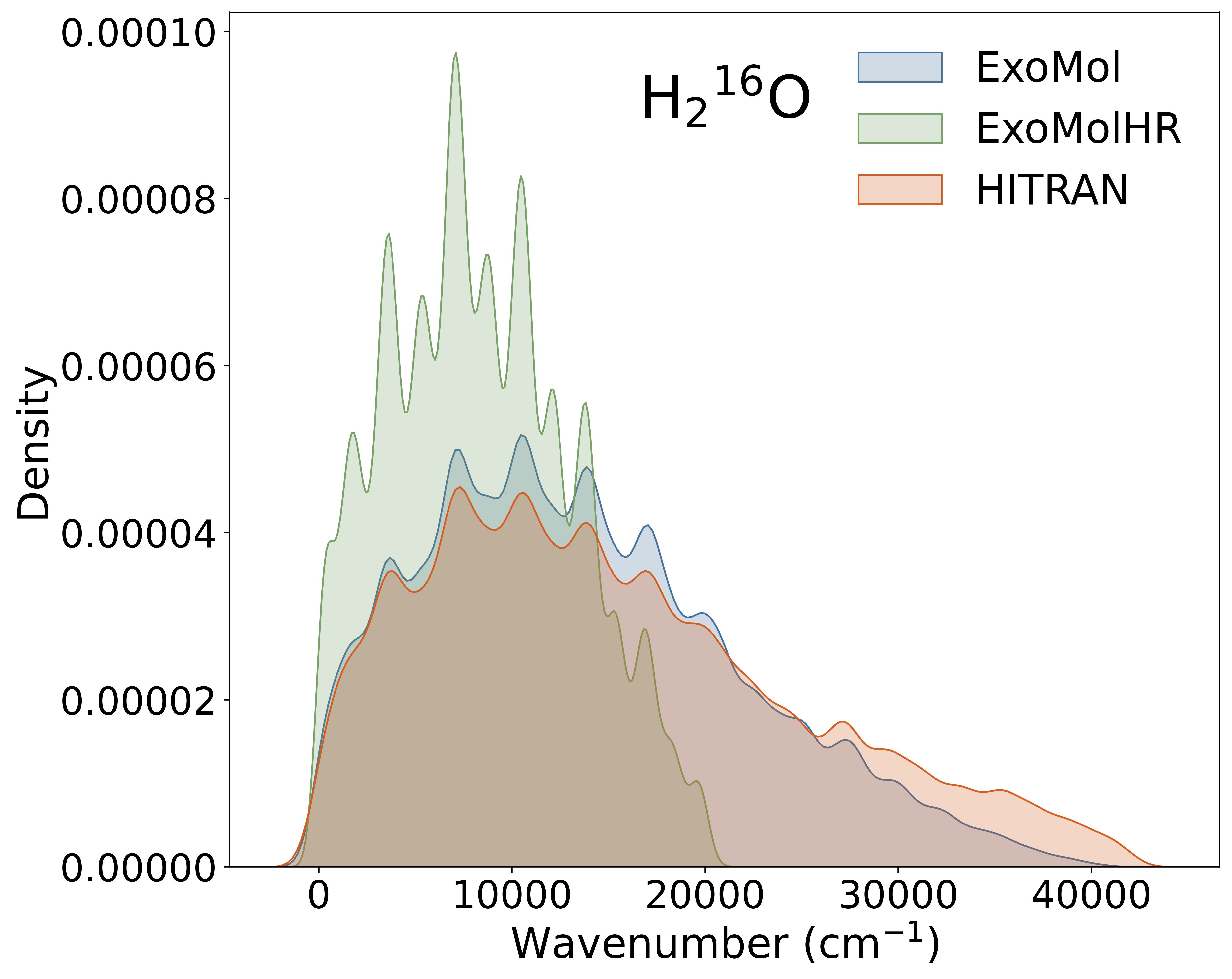}{0.33\textwidth}{}
          \fig{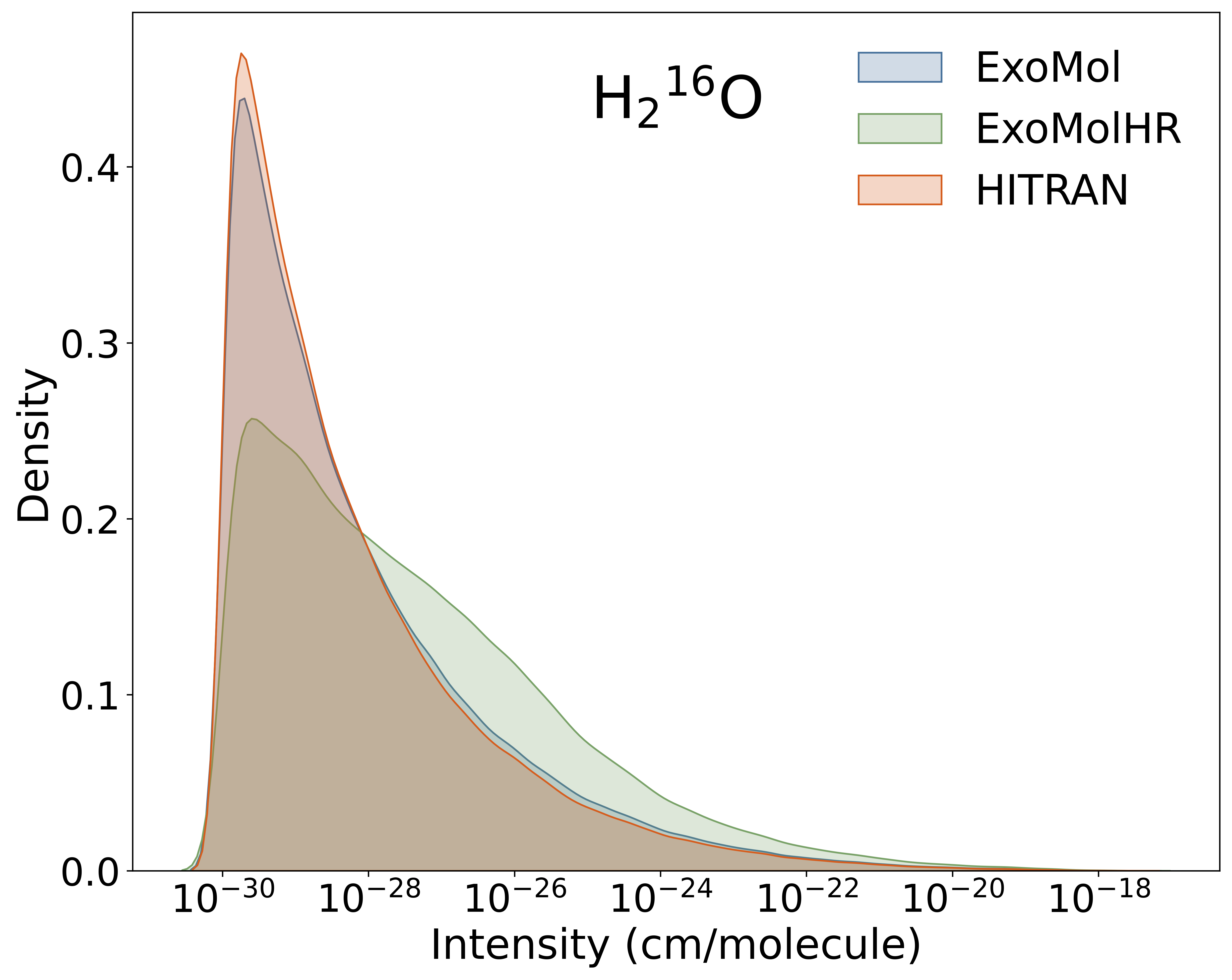}{0.33\textwidth}{}
          \fig{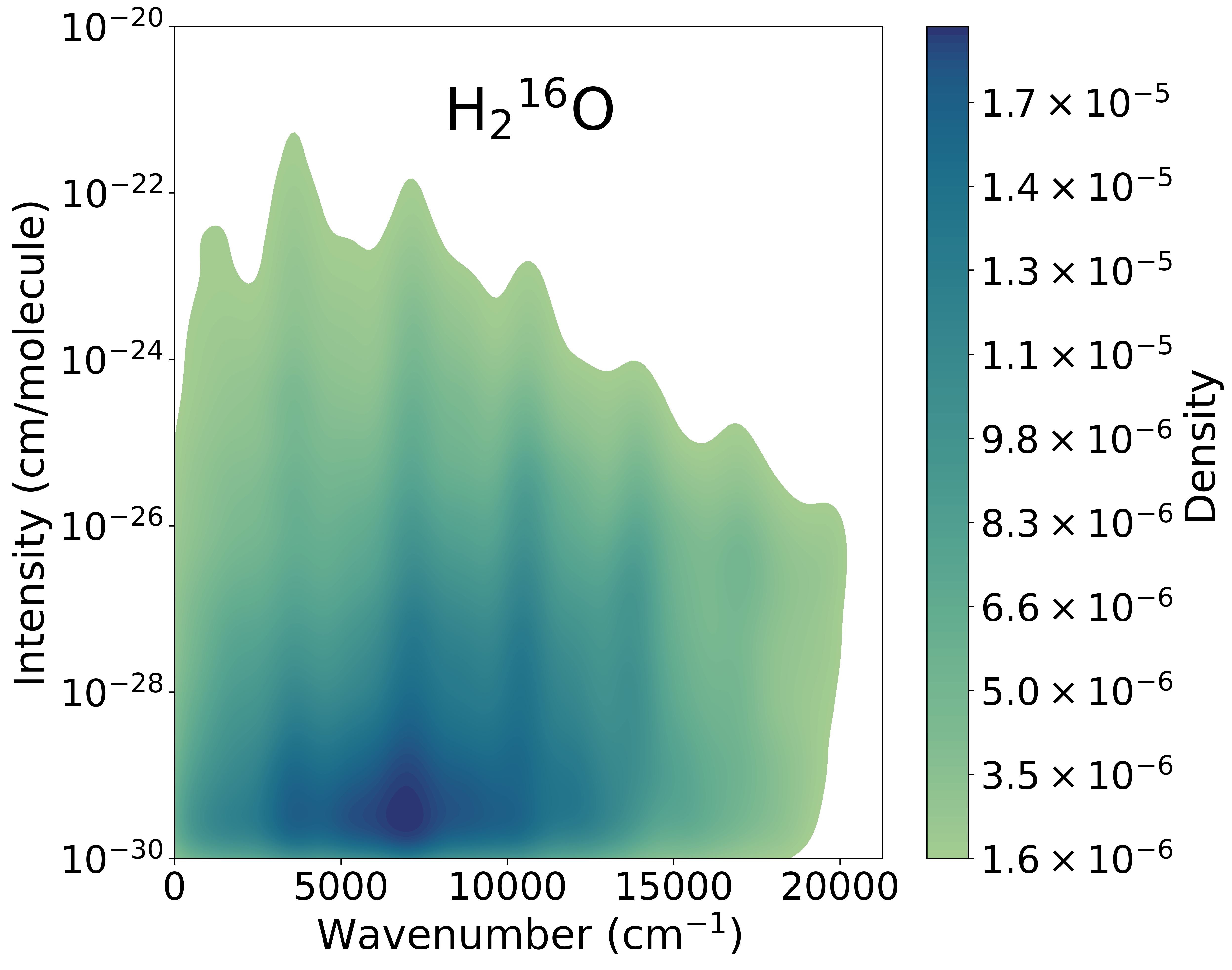}{0.33\textwidth}{}
          }
\gridline{\fig{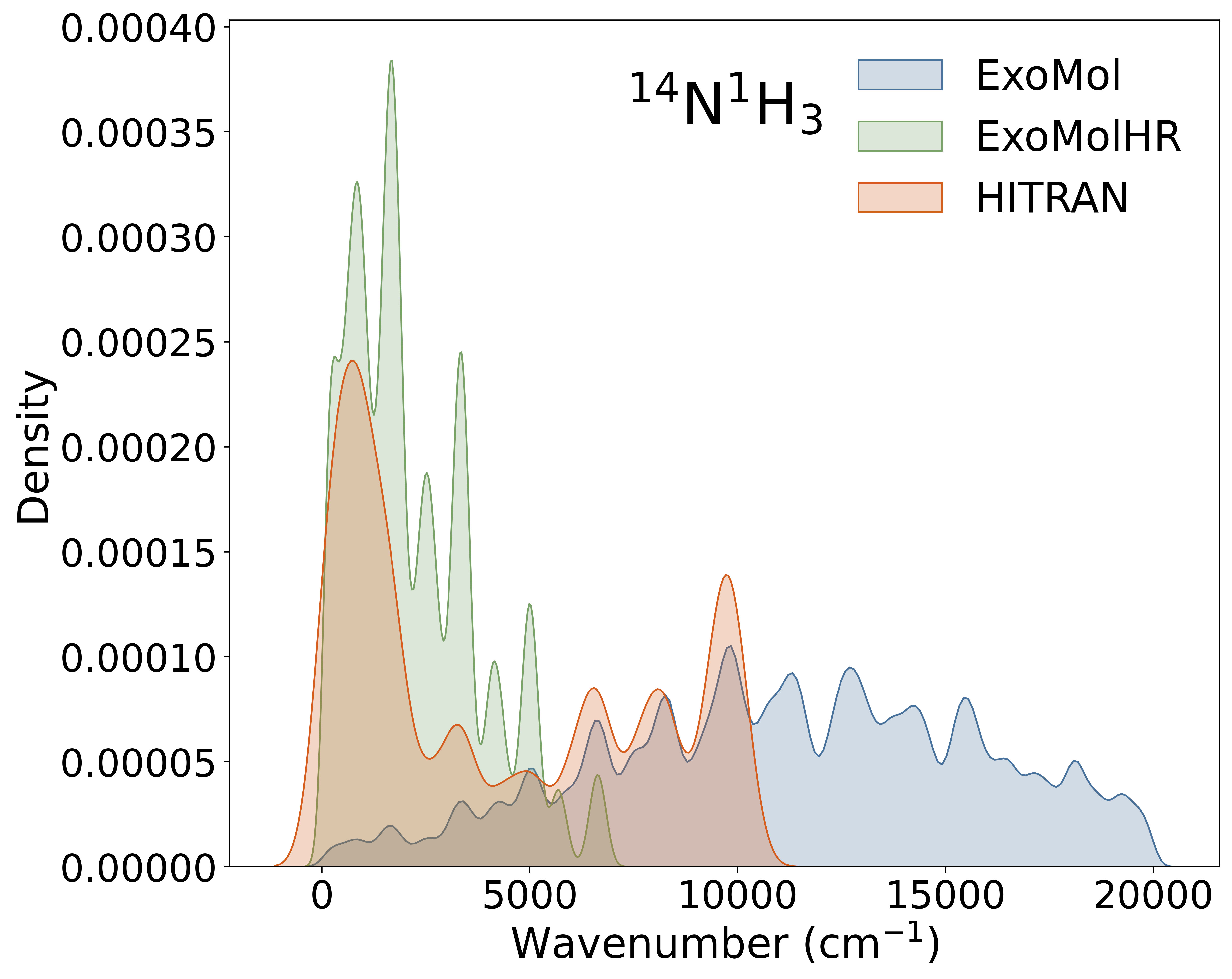}{0.33\textwidth}{}
          \fig{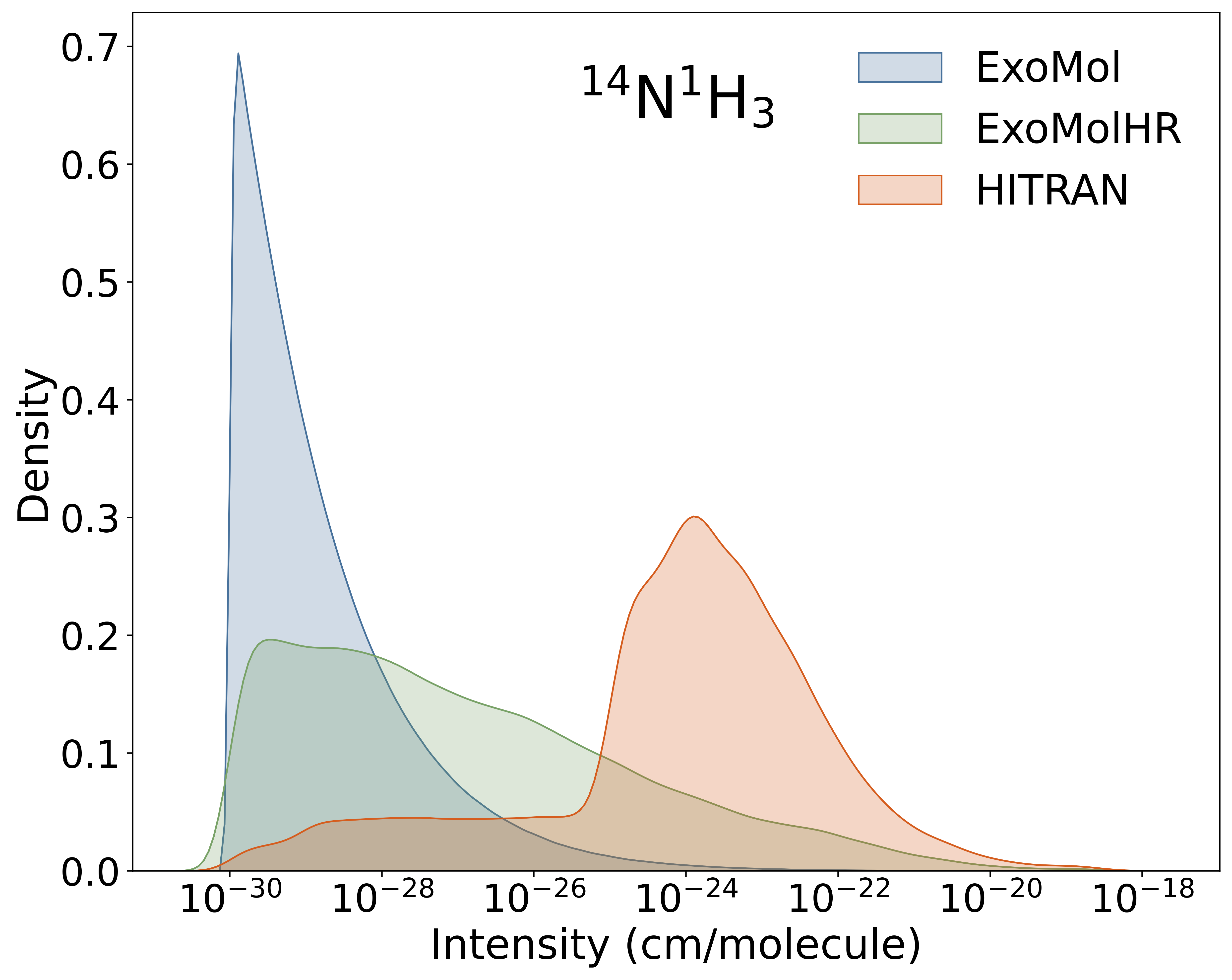}{0.33\textwidth}{}
          \fig{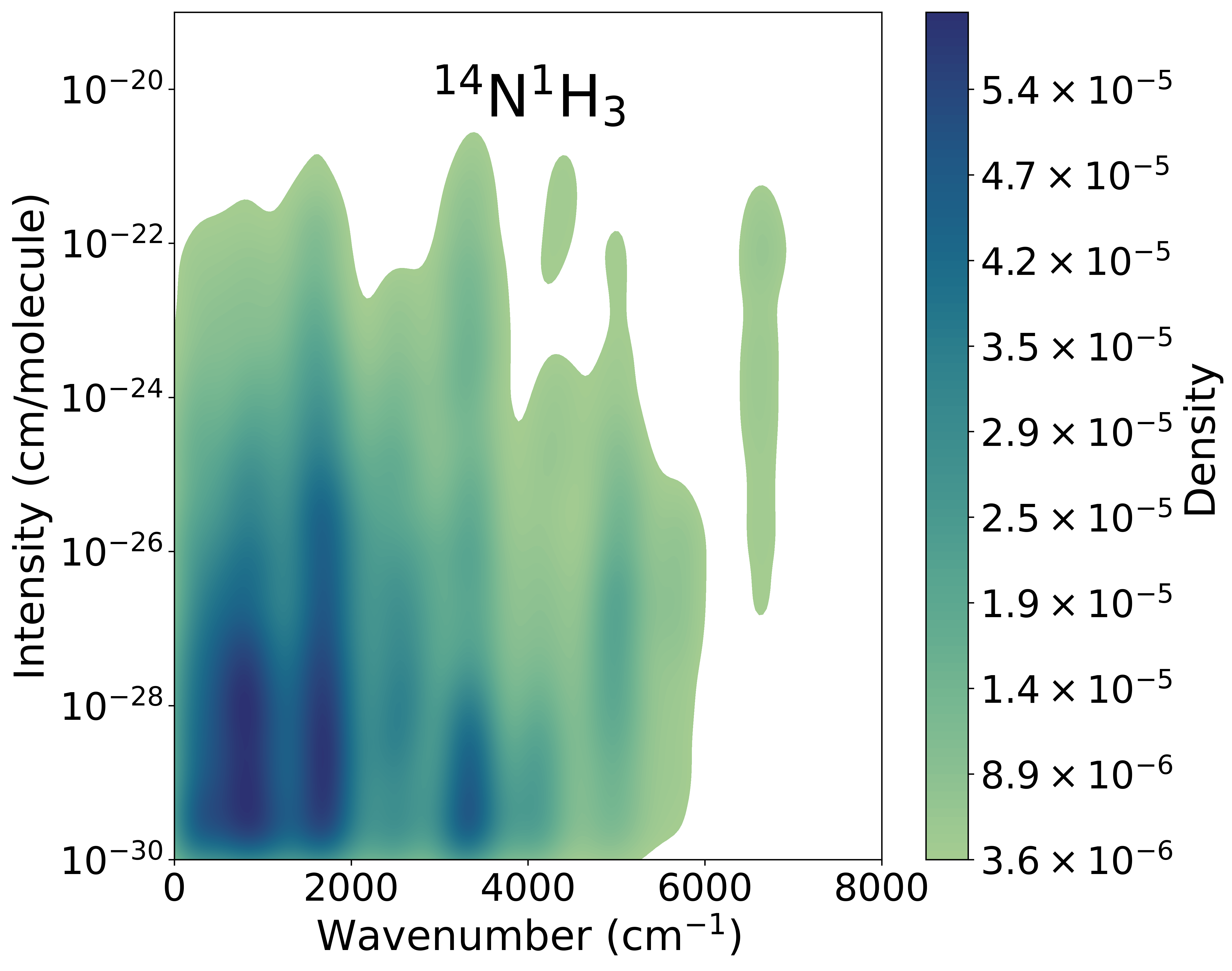}{0.33\textwidth}{}
          }
\gridline{\fig{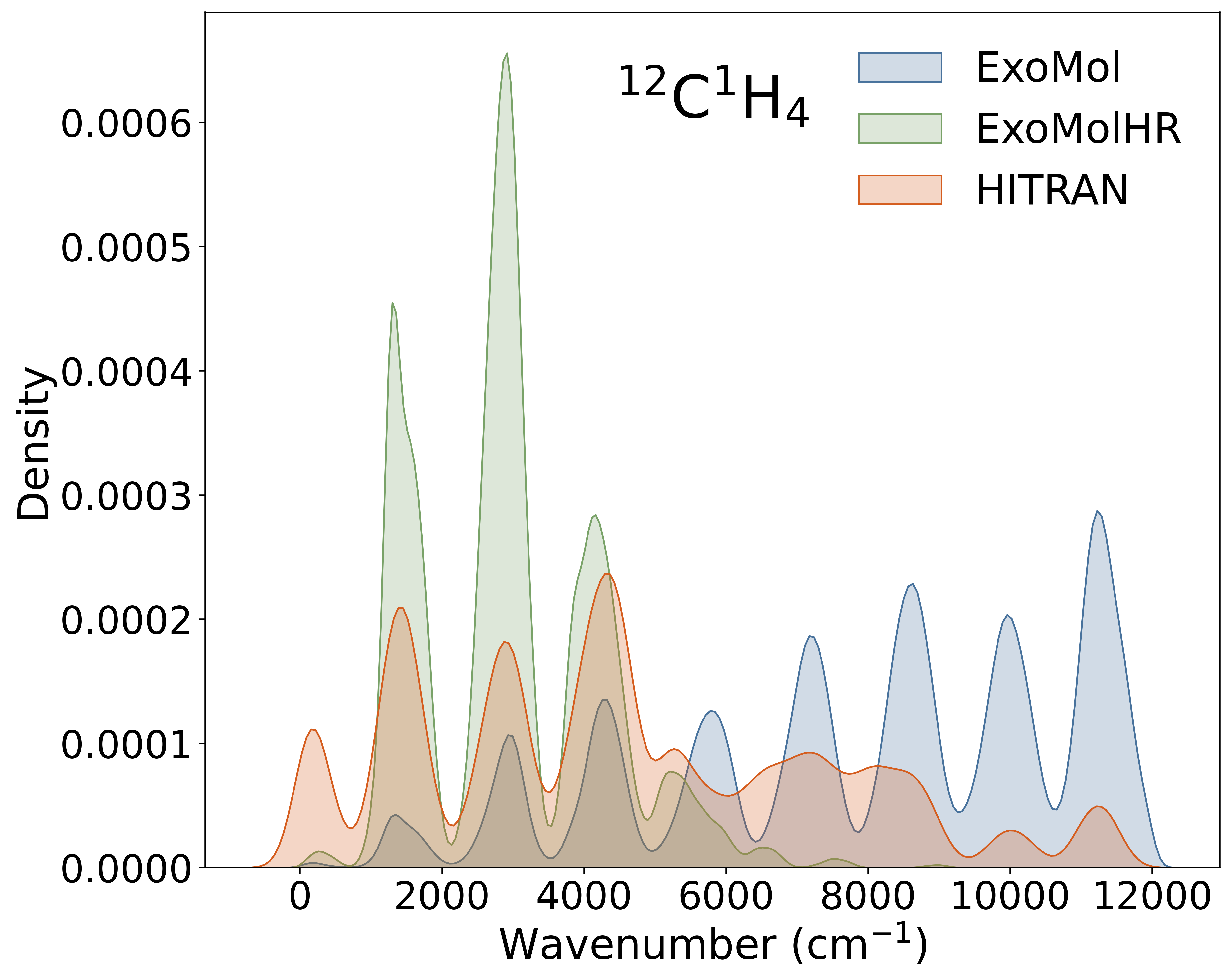}{0.33\textwidth}{}
          \fig{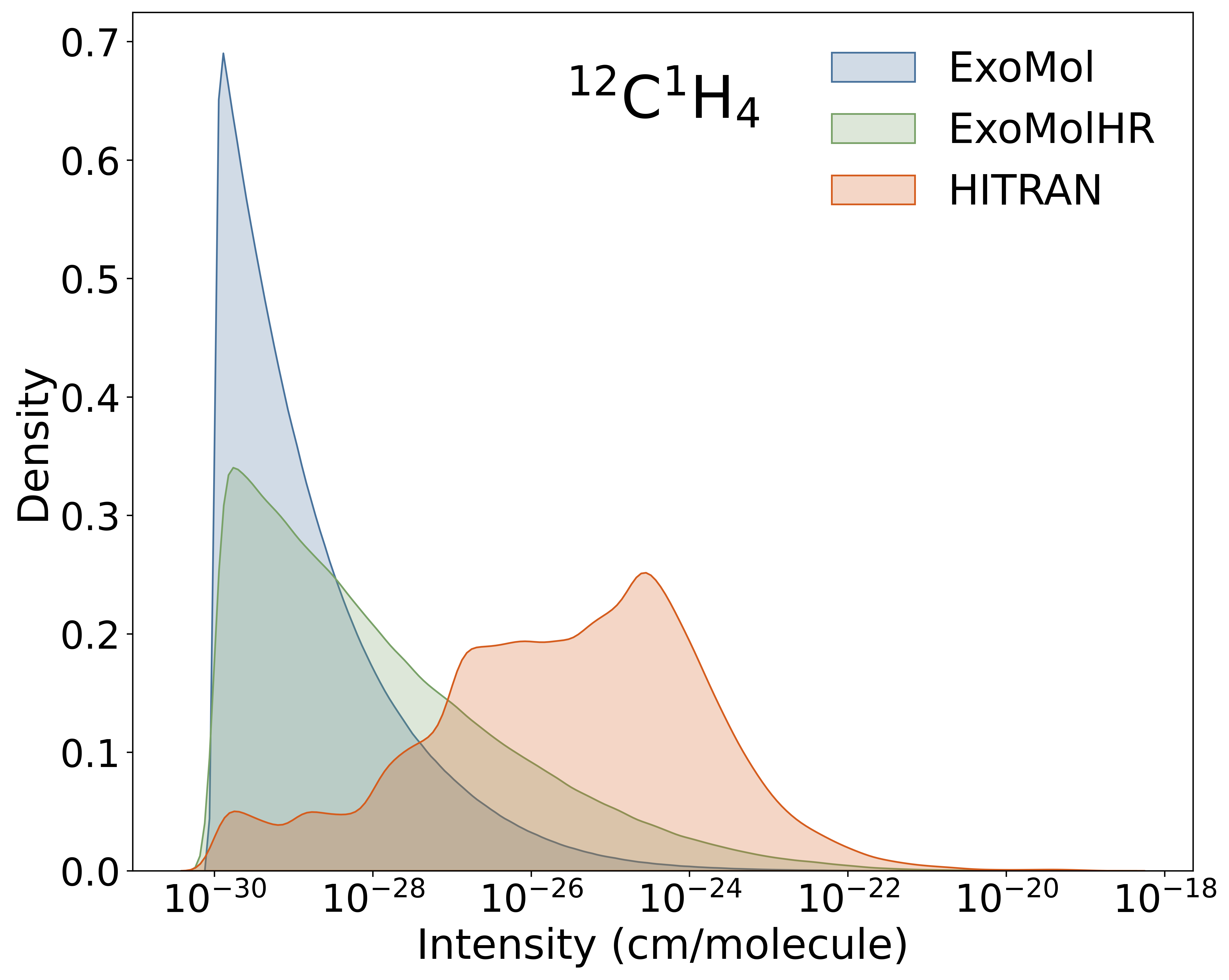}{0.33\textwidth}{}
          \fig{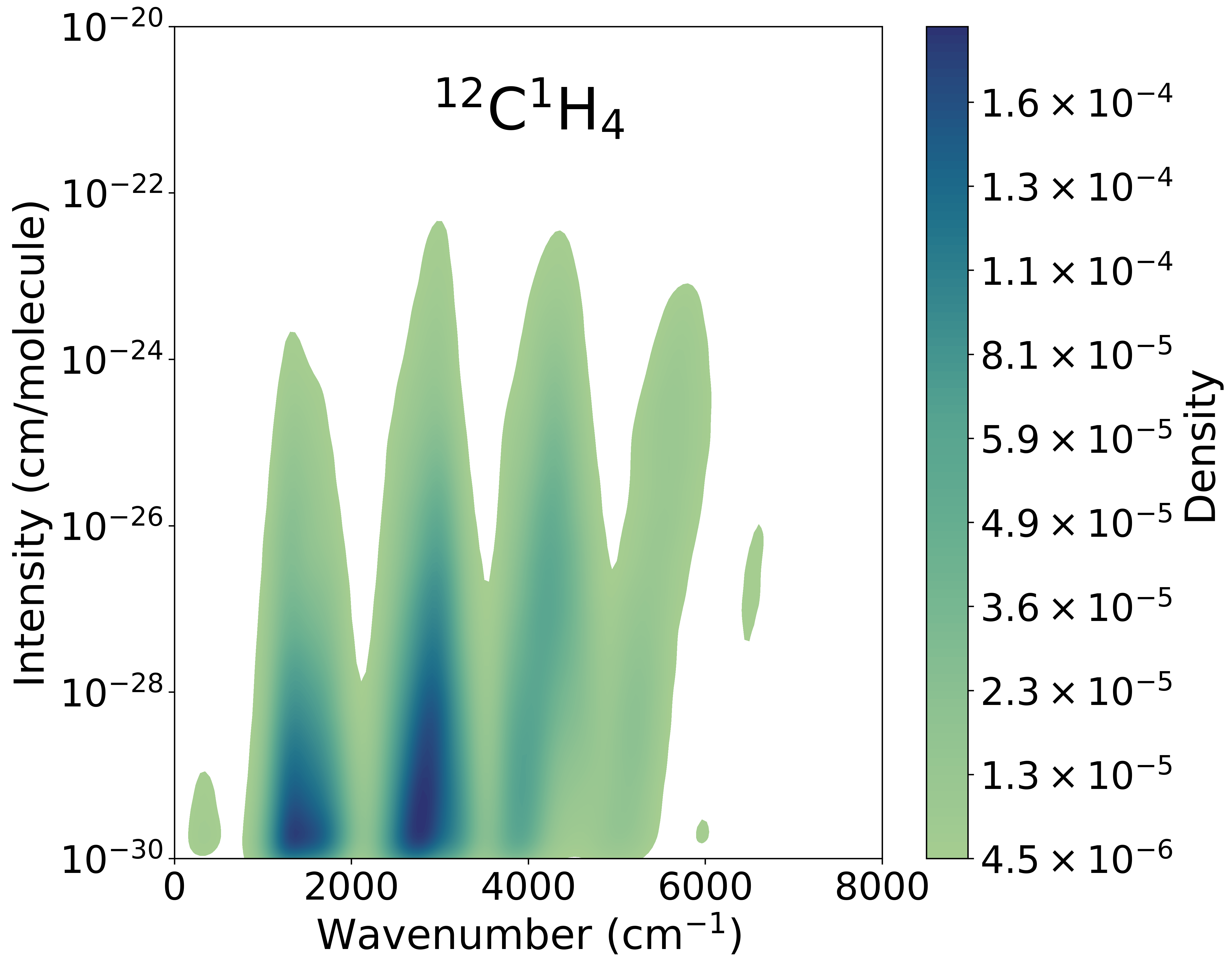}{0.33\textwidth}{}
          }
\gridline{\fig{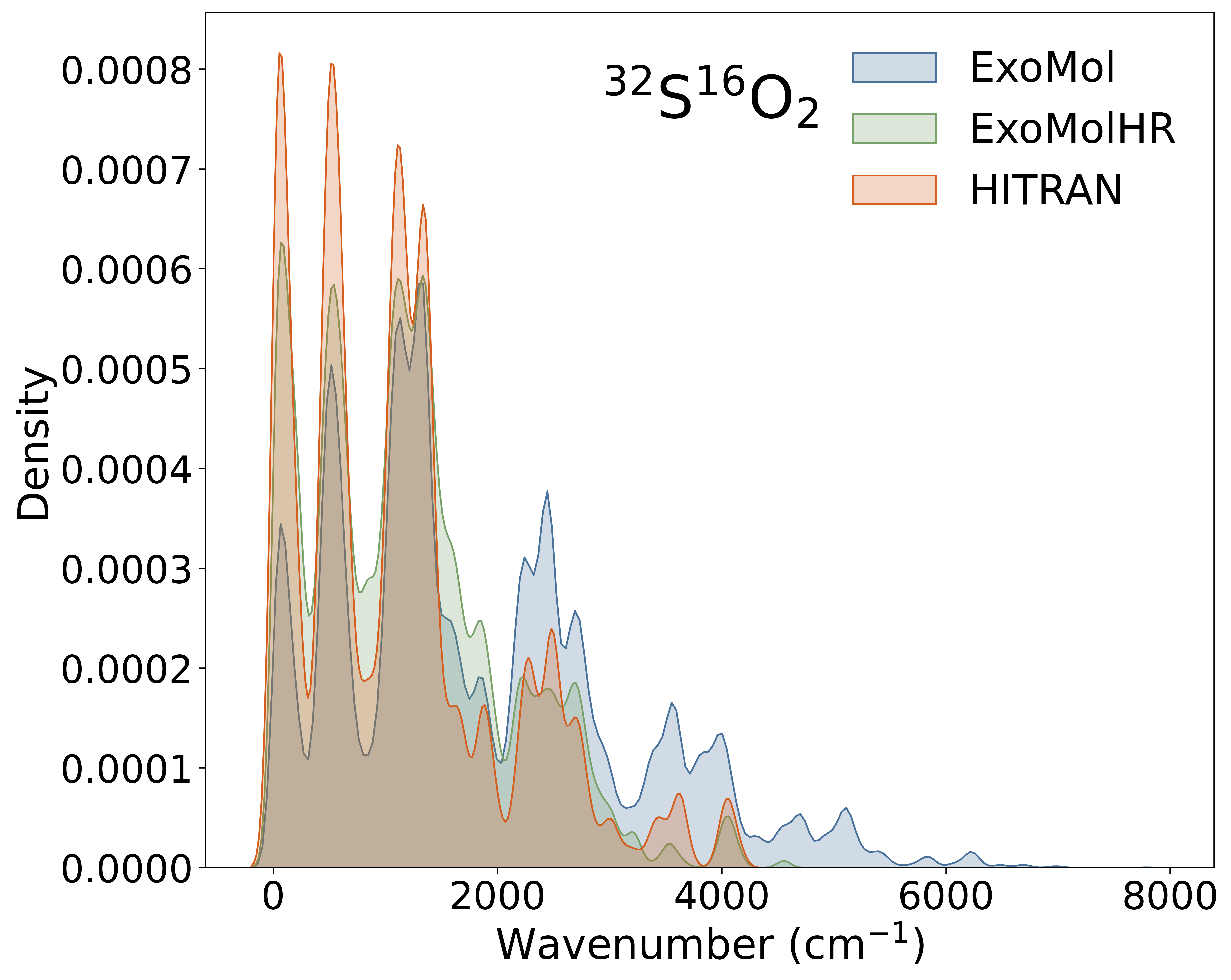}{0.33\textwidth}{}
          \fig{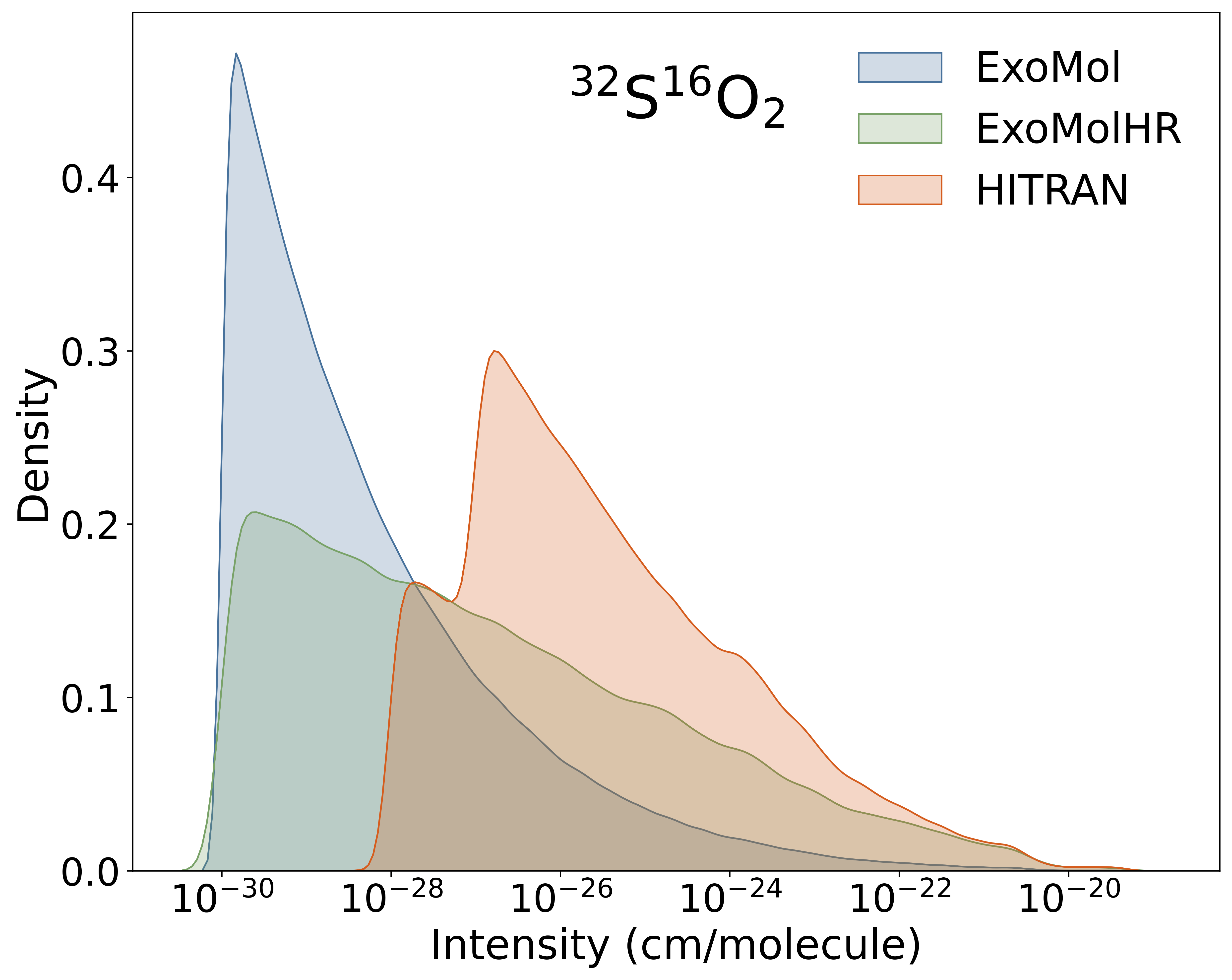}{0.33\textwidth}{}
          \fig{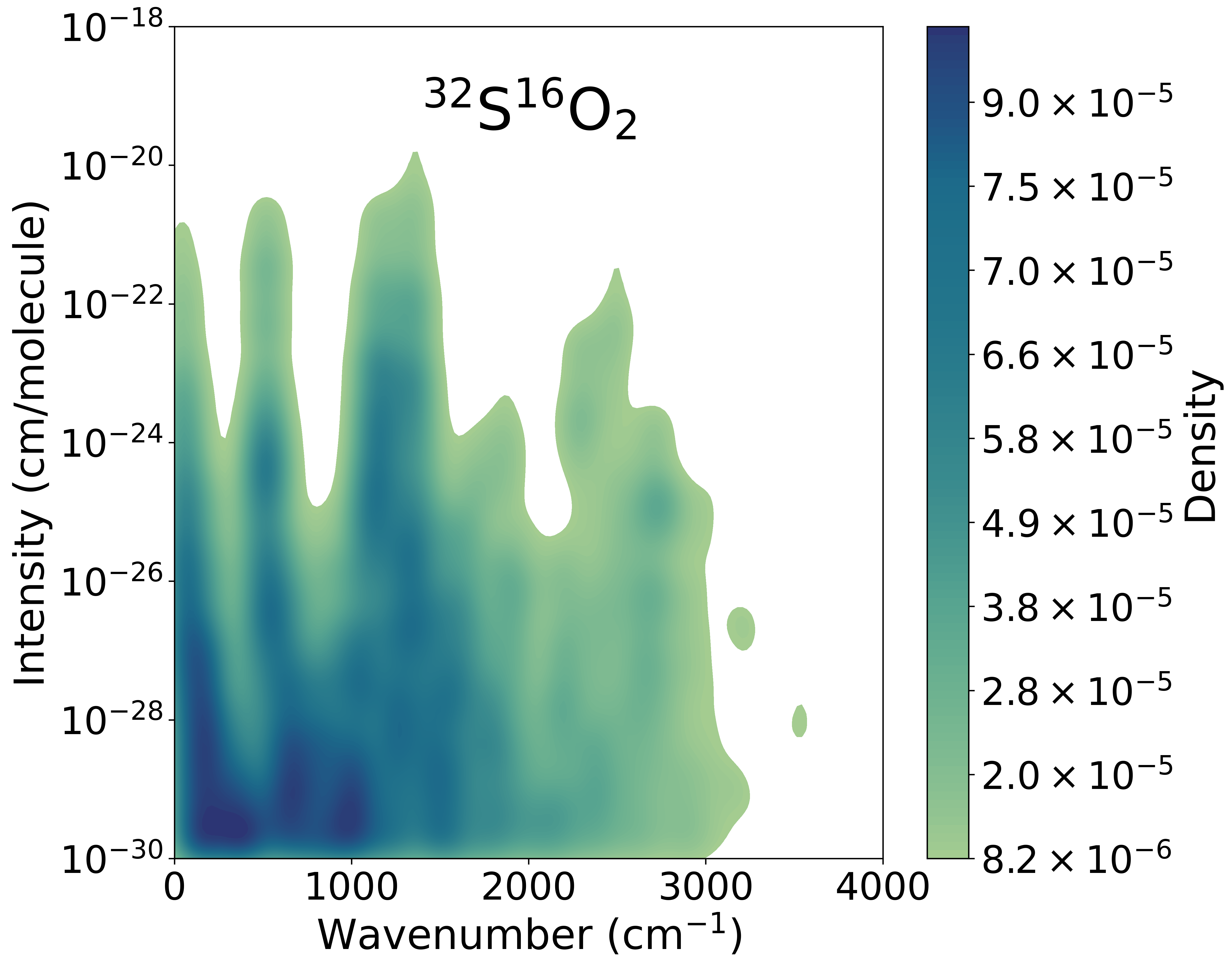}{0.33\textwidth}{}
          }
\caption{Compare the densities of wavenumber (left) and line intensities (middle) of the \textsc{ExoMol}, \ExoMolHR  and HITRAN  databases at $T=296$ K for $^1$H$_2$$^{16}$O, $^{14}$N$^1$H$_3$, $^{12}$C$^1$H$_4$, and $^{32}$S$^{16}$O$_2$. The \textsc{ExoMol} line lists provides full coverage using the POKAZATEL, CoYuTe, MM, and ExoAmes line list of $^1$H$_2$$^{16}$O, $^{14}$N$^1$H$_3$, $^{12}$C$^1$H$_4$, and $^{32}$S$^{16}$O$_2$, respectively. The 2D plots of the densities between wavenumbers and line intensities (cm$/$molecule) for \ExoMolHR databases at $T=296$ K are displayed on the right panel.\label{fig:densitycompare}}
\end{figure}

To analyze the distribution of the high-resolution line list among the range of the wavenumbers and intensities, we use $^1$H$_2$$^{16}$O, $^{14}$N$^1$H$_3$, $^{12}$C$^1$H$_4$, and $^{32}$S$^{16}$O$_2$ as samples in Figure~\ref{fig:densitycompare}. The left panel shows the densities for points in Figure~\ref{fig:3dbcompare} among the wavenumber ranges for the \textsc{ExoMol}, \textsc{ExoMolHR} and HITRAN databases. 
The left panel illustrates the relationship between the distribution density of spectral line points and the wavenumbers at room temperature. It is observed that compared to the comprehensive dataset provided by the \textsc{ExoMol} database, the spectra from high-resolution databases \ExoMolHR  and HITRAN  are predominantly concentrated in the lower wavenumber region.
The middle panel reveals that, after filtering out low-resolution and high-uncertainty line lists from the \textsc{ExoMol} database, there is a significant reduction in the density of low-intensity spectral lines for the \ExoMolHR database; we note that omitting such transitions can have a profound effect on the opacity \citep{jt572}.
The right panel provides a more intuitive representation of the density distribution of high-resolution spectra by depicting the relationship between wavenumber and intensity in the two-dimensional plots. These plots show that a higher concentration of points is present in the lower wavenumber range, lower intensity region. However, few high-resolution and higher-intensity spectra are distributed in the high wavenumber region.

\begin{longrotatetable}
\begin{deluxetable}{lllllcrrrrrrrl}
\tabletypesize{\scriptsize}
\setlength{\tabcolsep}{1.7mm}{
\tablecolumns{14} 
\tablecaption{Summary of lines with low uncertainties and high resolutions extracted from the \ExoMol database and used to form the \ExoMolHR database. \label{tab:HRlist}}
\tablehead{
\colhead{ID$_{\textrm{m}}$} & \colhead{Molecule} & \colhead{ID$_{\textrm{i}}$} & \colhead{Isotopologue} & \colhead{Dataset} & \colhead{Version} & \colhead{N$_{\textrm {states}}$} & \colhead{N$_{\textrm {files}}$} & \colhead{N$_{\textrm {trans}}$} & \colhead{Size$_{\textrm {trans}}$} & \colhead{N$_{\textrm {HRstates}}$} & \colhead{N$_{\textrm {HRlines}}$} & \colhead{Size$_{\textrm {HR}}$} & \colhead{\ExoMol line list}} 
\startdata
1  & AlCl     & 1  & $^{27}$Al$^{35}$Cl       & YNAT        & 20221231 & $65\,869$      & 1   & $4\,722\,048$       & 47.5 MB  & $41$      & $101$         & 15.3 KB  & \cite{jt887} \\
   &          & 2  & $^{27}$Al$^{37}$Cl       & YNAT        & 20221231 & $67\,507$      & 1   & $5\,748\,704$       & 57.7 MB  & $41$      & $121$         & 18.4 KB  & \\
2  & AlH      & 3  & $^{27}$Al$^1$H           & AloHa       & 20240307 & $1\,364$       & 1   & $29\,725$           & 152.9 KB & $135$     & $692$         & 104.2 KB & \cite{jt922} \\
3  & AlO      & 4  & $^{26}$Al$^{16}$O        & ATP         & 20210622 & $93\,350$      & 1   & $4\,866\,540$       & 53.6 MB  & $4\,783$  & $143\,197$    & 21.0 MB  & \cite{jt835} \\
   &          & 5  & $^{27}$Al$^{16}$O        & ATP         & 20210622 & $94\,862$      & 1   & $4\,945\,580$       & 31.1 MB  & $4\,980$  & $149\,577$    & 22.0 MB  & \\
   &          & 6  & $^{27}$Al$^{17}$O        & ATP         & 20210622 & $96\,350$      & 1   & $5\,148\,996$       & 56.7 MB  & $4\,787$  & $142\,905$    & 21.0 MB  & \\
   &          & 7  & $^{27}$Al$^{18}$O        & ATP         & 20210622 & $98\,269$      & 1   & $5\,365\,592$       & 59.1 MB  & $4\,799$  & $142\,976$    & 21.0 MB  & \\
4  & BeH      & 8  & $^9$Be$^1$H              & Darby-Lewis & 20240710 & $14\,950$      & 1   & $592\,308$          & 6.1 MB   & $132$     & $507$         & 76.4 KB  & \cite{jt792} \\
   &          & 9  & $^9$Be$^2$H              & Darby-Lewis & 20240710 & $14\,950$      & 1   & $689\,466$          & 7.2 MB   & $104$     & $310$         & 46.8 KB  & \\        
5  & C$_2$    & 10 & $^{12}$C$_2$             & 8states     & 20200628 & $44\,189$      & 1   & $6\,080\,920$       & 68.7 MB  & $8\,376$  & $445\,682$    & 65.5 MB  & \cite{jt809} \\
6  & C$_2$H$_2$ & 11 & $^{12}$C$_2$$^1$H$_2$  & aCeTY       & 20220918 & $5\,160\,803$  & 100 & $4\,347\,381\,911$  & 96.1 GB  & $8\,898$  & $473\,850$    & 47.9 MB  & \cite{jt780} \\
7  & CH$_4$   & 12 & $^{12}$C$^1$H$_4$        & MM          & 20240113 & $9\,155\,208$  & 121 & $50\,395\,644\,806$ & 439.0 GB & $21\,021$ & $7\,649\,736$ & 760.2 MB & \cite{jt927} \\
8  & CN       & 13 & $^{12}$C$^{14}$N         & Trihybrid   & 20210526 & $28\,004$      & 1   & $2\,285\,103$       & 25.8 MB  & $4\,833$  & $244\,808$    & 36.0 MB  & \cite{21SyMcXx.CN} \\
9  & CO$_2$   & 14 & $^{12}$C$^{16}$O$_2$     & UCL-4000    & 20200630 & $3\,480\,477$  & 20  & $2\,557\,549\,946$  & 21.7 GB  & $18\,881$ & $2\,600\,218$ & 362.0 MB & \cite{jt804} \\
10 & CaH      & 15 & $^{40}$Ca$^1$H           & XAB         & 20220211 & $6\,825$       & 1   & $293\,151$          & 2.3 MB   & $1\,165$  & $12\,341$     & 1.8 MB   & \cite{jt858} \\
11 & CaOH     & 16 & $^{40}$Ca$^{16}$O$^1$H   & OYT6        & 20230523 & $3\,187\,52$2  & 18  & $23\,384\,729\,495$ & 202.0 GB & $1\,424$  & $12\,984$     & 1.6 MB   & \cite{jt878} \\
12 & H$_2$CO  & 17 & $^1$H$_2$$^{12}$C$^{16}$O & AYTY       & 20230430 & $10\,297\,025$ & 100 & $12\,688\,112\,669$ & 111.1 GB & $4\,813$  & $317\,729$    & 41.2 MB  & \cite{jt597} \\
13 & H$_2$CS  & 18 & $^1$H$_2$$^{12}$C$^{32}$S & MOTY       & 20221231 & $52\,292\,454$ & 8   & $43\,561\,116\,660$ & 399.5 GB & $3\,625$  & $72\,218$     & 6.6 MB   & \cite{jt886} \\
14 & H$_2$O   & 19 & $^1$H$_2$$^{16}$O        & POKAZATEL   & 20230621 & $810\,269$     & 412 & $5\,745\,071\,340$  & 48.1 GB  & $14\,395$ & $3\,520\,554$ & 396.2 MB & \cite{jt734} \\
15 & H$_2$S   & 20 & $^1$H$_2$$^{32}$S        & AYT2        & 20220918 & $220\,631$     & 35  & $115\,032\,941$     & 873.0 MB & $7\,314$  & $1\,084\,520$ & 122.0 MB & \cite{jt640} \\
16 & H$_3$O$^+$ & 21 & $^1$H$_3$$^{16}$O$^+$  & eXeL        & 20200707 & $1\,173\,114$  & 100 & $2\,089\,331\,073$  & 10.1 GB  & $232$     & $1\,785$      & 251.3 KB & \cite{jt805} \\
17 & H$_3$$^+$  & 22 & $^1$H$_2$$^2$H$^+$     & ST          & 20230123 & $33\,330$      & 1   & $22\,164\,810$      & 128.2 MB & $109$     & $646$         & 87.2 KB  & \cite{jt890} \\
   &          & 23 & $^1$H$_3$$^+$            & MiZATeP     & 20230123 & $158\,721$     & 1   & $127\,542\,657$     & 1.7 GB   & $994$     & $13\,606$     & 1.9 MB   & \\
   &          & 24 & $^2$H$_2$$^1$H$^+$       & MiZo        & 20221221 & $369\,500$     & 32  & $2\,290\,235\,000$  & 14.8 GB  & $115$     & $683$         & 92.2 KB  & \\
   &          & 25 & $^2$H$_3$$^+$            & MiZo        & 20230124 & $37\,410$      & 21  & $36\,078\,183$      & 170.3 MB & $115$     & $225$         & 32.0 KB  & \\
18 & LiOH     & 26 & $^6$Li$^{16}$O$^1$H      & OYT7        & 20231001 & $192\,412$     & 5   & $294\,573\,305$     & 2.3 GB   & $255$     & $840$         & 113.3 KB & \cite{jt905} \\
   &          & 27 & $^7$Li$^{16}$O$^1$H      & OYT7        & 20230529 & $203\,762$     & 5   & $331\,274\,717$     & 2.6 GB   & $240$     & $749$         & 101.1 KB & \\
19 & MgH      & 28 & $^{24}$Mg$^1$H           & XAB         & 20220211 & $3\,148$       & 1   & $88\,575$           & 953.2 KB & $237$     & $2\,462$      & 370.4 KB & \cite{jt858} \\
   &          & 29 & $^{25}$Mg$^1$H           & XAB         & 20220211 & $3\,156$       & 1   & $88\,776$           & 955.6 KB & $548$     & $5\,850$      & 879.9 KB & \\
   &          & 30 & $^{26}$Mg$^1$H           & XAB         & 20220211 & $3\,160$       & 1   & $88\,891$           & 956.5 KB & $537$     & $5\,339$      & 803.1 KB & \\
20 & N$_2$O   & 31 & $^{14}$N$_2$$^{16}$O     & TYM         & 20240527 & $1\,759\,068$  & 11  & $1\,360\,351\,722$  & 11.0 GB  & $17\,018$ & $3\,459\,640$ & 309.9 MB & \cite{jt951} \\
21 & NH       & 32 & $^{14}$N$^1$H            & kNigHt      & 20240301 & $4\,076$       & 1   & $327\,014$          & 1.6 MB   & $1\,030$  & $26\,131$     & 3.7 MB   & \cite{24PeMcxx.NH} \\
   &          & 33 & $^{14}$N$^2$H            & kNigHt      & 20240301 & $7\,406$       & 1   & $778\,105$          & 3.7 MB   & $118$     & $943$         & 136.4 KB & \\
   &          & 34 & $^{15}$N$^1$H            & kNigHt      & 20240301 & $4\,089$       & 1   & $327\,877$          & 1.6 MB   & $118$     & $943$         & 136.4 KB & \\
   &          & 35 & $^{15}$N$^2$H            & kNigHt      & 20240301 & $7\,465$       & 1   & $785\,940$          & 3.8 MB   & $118$     & $943$         & 136.4 KB & \\
22 & NH$_3$   & 36 & $^{14}$N$^1$H$_3$        & CoYuTe      & 20200730 & $5\,095\,730$  & 200 & $16\,941\,637\,250$ & 142.9 GB & $4\,720$  & $412\,149$    & 64.4 MB  & \cite{jt771} \\
   &          & 37 & $^{15}$N$^1$H$_3$        & CoYuTe-15   & 20240808 & $12\,699\,617$ & 10  & $929\,795\,249$     & 7.3 GB   & $2\,699$  & $148\,651$    & 22.0 MB & \cite{jt952} \\
23 & NO       & 38 & $^{14}$N$^{16}$O         & XABC        & 20210422 & $30\,811$      & 1   & $4\,596\,666$       & 96.7 MB  & $3\,044$  & $106\,711$    & 15.7 MB  & \cite{jt831} \\
24 & OCS      & 39 & $^{16}$O$^{12}$C$^{32}$S & OYT8        & 20240425 & $2\,399\,110$  & 10  & $2\,527\,364\,150$  & 26.4 GB  & $5\,198$  & $279\,273$    & 44.7 MB  & \cite{jt943} \\
25 & PN       & 40 & $^{31}$P$^{14}$N         & PaiN        & 20240505 & $30\,327$      & 1   & $1\,333\,445$       & 15.6 MB  & $32$      & $44$          & 6.8 KB   & \cite{jt954} \\
26 & SO       & 41 & $^{32}$S$^{16}$O         & SOLIS       & 20230914 & $84\,114$      & 1   & $7\,086\,100$       & 81.6 MB  & $536$     & $2\,501$      & 366.5 KB & \cite{jt924} \\
27 & SO$_2$   & 42 & $^{32}$S$^{16}$O$_2$     & ExoAmes     & 20170131 & $3\,270\,271$  & 80  & $1\,300\,000\,000$  & 25.4 GB  & $14\,924$ & $1\,504\,495$ & 163.6 MB & \cite{jt635} \\
28 & SiN      & 43 & $^{28}$Si$^{14}$N        & SiNfull     & 20220809 & $131\,936$     & 1   & $43\,646\,806$      & 506.9 MB & $99$      & $670$         & 100.9 KB & \cite{jt875} \\
   &          & 44 & $^{28}$Si$^{15}$N        & SiNfull     & 20220809 & $133\,460$     & 1   & $44\,816\,182$      & 520.9 MB & $56$      & $464$         & 69.9 KB  & \\
   &          & 45 & $^{29}$Si$^{14}$N        & SiNfull     & 20220809 & $132\,335$     & 1   & $43\,946\,969$      & 510.6 MB & $56$      & $464$         & 69.9 KB  & \\
   &          & 46 & $^{30}$Si$^{14}$N        & SiNfull     & 20220809 & $132\,706$     & 1   & $44\,223\,730$      & 513.9 MB & $564$      & $464$         & 69.9 KB  & \\
29 & SiO      & 47 & $^{28}$Si$^{16}$O        & SiOUVenIR   & 20211105 & $174\,250$     & 1   & $91\,395\,763$      & 1.1 GB   & $911$     & $8\,729$      & 1.3 MB   & \cite{jt847} \\
30 & TiO      & 48 & $^{48}$Ti$^{16}$O        & Toto        & 20210825 & $301\,245$     & 1   & $58\,983\,952$      & 689.5 MB & $8\,725$  & $499\,775$    & 73.4 MB  & \cite{jt760}\\
31 & VO       & 49 & $^{51}$V$^{16}$O         & HyVO        & 20231218 & $3\,410\,598$  & 90  & $58\,904\,173\,243$ & 297.9 GB & $7\,043$  & $635\,722$    & 68.1 MB  & \cite{jt923} \\
32 & YO       & 50 & $^{89}$Y$^{16}$O         & BRYTS       & 20230916 & $173\,621$     & 1   & $60\,678\,140$      & 719.1 MB & $28$      & $25$          & 3.9 KB   & \cite{jt921} \\
   &          & 51 & $^{89}$Y$^{17}$O         & BRYTS       & 20230916 & $182\,598$     & 1   & $62\,448\,157$      & 740.9 MB & $28$      & $25$          & 3.9 KB   & \\
   &          & 52 & $^{89}$Y$^{18}$O         & BRYTS       & 20230916 & $182\,547$     & 1   & $64\,164\,605$      & 761.9 MB & $28$      & $25$          & 3.9 KB   & \\
33 & ZrO      & 53 & $^{90}$Zr$^{16}$O        & ZorrO       & 20230713 & $227\,118$     & 1   & $47\,662\,773$      & 369.3 MB & $5\,313$  & $145\,317$    & 21.3 MB  & \cite{23PeTaMc.ZrO} \\
   &          & 54 & $^{91}$Zr$^{16}$O        & ZorrO       & 20230713 & $227\,118$     & 1   & $47\,748\,501$      & 370.1 MB & $1\,058$  & $5\,164$      & 776.8 KB & \\
   &          & 55 & $^{92}$Zr$^{16}$O        & ZorrO       & 20230713 & $227\,124$     & 1   & $47\,830\,250$      & 370.8 MB & $1\,058$  & $5\,164$      & 776.8 KB & \\
   &          & 56 & $^{93}$Zr$^{16}$O        & ZorrO       & 20230713 & $227\,126$     & 1   & $47\,928\,979$      & 371.6 MB & $1\,058$  & $5\,164$      & 776.8 KB & \\
   &          & 57 & $^{94}$Zr$^{16}$O        & ZorrO       & 20230713 & $227\,128$     & 1   & $47\,994\,352$      & 372.1 MB & $1\,058$  & $5\,164$      & 776.8 KB & \\
   &          & 58 & $^{96}$Zr$^{16}$O        & ZorrO       & 20230713 & $227\,134$     & 1   & $48\,136\,388$      & 373.3 MB & $1\,058$  & $5\,164$      & 776.8 KB & \\
\enddata
}
\tablecomments{\\
ID$_{\textrm{m}}$: Molecule index; \\
ID$_{\textrm{i}}$: Isotopologue index; \\
Verison: \ExoMol dataset update date in the format \texttt{YYYYMMDD}; \\
N$_{\textrm {states}}$: Number of states in states file; \\
N$_{\textrm {files}}$: Number of transitions files; \\
N$_{\textrm {trans}}$: Number of transitions in transitions file(s); \\
Size$_{\textrm {trans}}$: Total file size of bzip2 compressed transitions file(s); \\
N$_{\textrm {HRstates}}$: Number of states with uncertainties $\Delta E \le 0.01$ cm$^{-1}$; \\
N$_{\textrm {HRlines}}$: Number of lines in \ExoMolHR dataset with $R>100\,000$; \\
Size$_{\textrm {HR}}$: File size of uncompressed \ExoMolHR dataset; \\
\ExoMol line list: Data source reference.
}
\end{deluxetable}
\end{longrotatetable}


\section{Conclusions} \label{sec:conclusions}

\ExoMolHR is a new high-resolution molecular spectroscopic database, developed from the \textsc{ExoMol} database which is designed for high-temperature molecular line lists for modelling exoplanet atmospheres. 
The
\ExoMolHR database is based on resolving power calculated  by filtering out states whose energies are only known with high uncertainty; transition wavenumber/wavelengths known to better than 1 part in $100\,000$  are subsequently extracted to give high-resolution line lists.
The \ExoMolHR database contains \nhrl transitions associated with \nhrs states which have been determined with low uncertainties ($\Delta \nu \le$ 0.01 cm$^{-1}$) from  \nexomolt transitions and \nexomols states for the same datasets in the \ExoMol database.
We note that our procedure will inevitably lead to missing lines
in the \ExoMolHR compilation. Users who wish to check for missing lines or to looking for completeness
should use  the original \textsc{ExoMol} line lists with one of the spectral generator programs \textsc{ExoCross}
\citep{jt708} or \textsc{PyExoCross} \citep{jt914}.

The \ExoMolHR website allows users to select molecules and isotopologues, calculate intensities based on user-specified temperatures, and apply filters with the desired wavenumber range and intensity threshold. The website provides a download link for a compressed folder containing the high-resolution line lists in \texttt{.csv} format for each isotopologue, along with the total folder size and the number of whole line lists. Additionally, the site provides a  spectral viewer which generates stick spectra for all selected isotopologues within the specified temperature, wavenumber range, and intensity threshold. The webpage also offers interactive zooming and panning functionality for the spectral plot. The \ExoMolHR database is freely accessible through an interactive web interface \url{https://www.exomol.com/exomolhr/}. 
An API has been designed to facilitate user data requests, with the option to dynamically access and view data directly on the website.

The molecular data in the \ExoMolHR database are derived from the \textsc{ExoMol} database, with the low-uncertainty line lists computed using data from the MARVEL database.
At the time of writing, the \ExoMolHR database supports \niso isotopologues from \nmol molecules and this number is increasing  as additional MARVEL studies are completed.


\section*{Acknowledgments}
This work was supported by the European Research Council (ERC)  under Advanced Investigator Project 883830 and by UK STFC under grant ST/Y001508/1.

\section*{Data availability statement}
The data that support the findings of this study are openly available at the following URL: \url{https://www.exomol.com/exomolhr/}. 
The molecular line lists sources discussed in this paper are available from the \textsc{ExoMol} website \url{https://www.exomol.com/}. 
The \textsc{PyExoCross} program used for calculating in this study is publicly accessible on GitHub \href{https://github.com/ExoMol/PyExoCross.git}{https://github.com/ExoMol/PyExoCross.git} and its program manual webpage is \url{https://pyexocross.readthedocs.io/}.

%


\software{\textsc{PyExoCross} \citep{jt914}}





\bibliographystyle{aasjournal}



\end{document}